\documentclass[a4paper, 11pt]{article}
\usepackage{jcappub,cancel}
\usepackage{paralist}
\usepackage{booktabs}
\usepackage{aas_macros}
\usepackage{subcaption}
\usepackage[a4paper]{geometry}
\graphicspath{{images/}}
\title{No WIMP Mini-Spikes in Dwarf Spheroidal Galaxies}

\author[a]{Mark Wanders,}
\author[a]{Gianfranco Bertone,}
\author[b]{Marta Volonteri,}
\author[a]{and Christoph Weniger}
\affiliation[a]{GRAPPA, University of Amsterdam,\\ Science Park 904, Amsterdam, the Netherlands}
\affiliation[b]{Institut d'Astrophysique de Paris,\\ 98bis Boulevard Arago, Paris, France}

\newcommand{\Fermi}{\textit{Fermi}}

\begin{document}

\abstract{The formation of black holes inevitably affects the distribution of
dark and baryonic matter in their vicinity, leading to an enhancement of the
dark matter density, called {\it spike}, and if dark matter is made of WIMPs, 
to a strong enhancement of the dark matter annihilation rate. Spikes
at the center of galaxies like the Milky Way are efficiently disrupted by
baryonic processes, but {\it mini-spikes}
can form and survive undisturbed at the center of dwarf spheroidal galaxies. 
We show that \Fermi\ LAT satellite data allow to set very stringent
limits on the existence of mini-spikes in dwarf galaxies: for thermal WIMPs with mass between 
100 GeV and 1 TeV, we obtain a maximum black hole mass between 100 and 1000 $M_\odot$,
ruling out black holes masses extrapolated from the M-$\sigma$ relationship in a large 
region of the parameter space. 
We also performed Monte Carlo simulations of merger histories of black holes in dwarf
spheroidals in a scenario where black holes form from the direct collapse of
primordial gas in early halos, and found that this specific formation scenario is 
incompatible at the $84\%$ CL with dark matter being in the form of thermal WIMPs.}

\maketitle

\section{Introduction}

In most scenarios that lead to the formation of supermassive
black holes (SMBHs, $M_\bullet \sim 10^6$--$10^8 M_{\odot}$), the growth of
these objects from a small seed inevitably affects the distribution of baryonic
matter around them (e.g.~refs.~\cite{peebles1972, Quinlan:1994ed}) and can lead
to large overdensities in the dark matter (DM) \cite{BertoneBook} distribution
called {\it spikes} \citep{gondolosilk1999}.  If DM is made of Weakly
Interacting Massive Particles (WIMPs) \cite{Jungman:1995df,
Bergstrom:2000pn,bhs2005}, these overdensities imply a strong enhancement of
the DM annihilation rate at the Galactic center \cite{gondolosilk1999,
Gondolo:2000pn, Bertone:2001jv, Bertone:2002je} and more in general around any
SMBH, boosting the predicted extragalactic gamma-ray background
\cite{Ahn:2007ty, Belikov:2013nca}. However, mergers, off-center formation of
the seed black holes (BHs) and gravitational scattering off of stars are likely
to disrupt spikes at the center of galaxies like the Milky Way (MW)
\citep{uzk2001, merritt2002, bertonemerrit2005}. 

These overdensities would instead persist around intermediate mass black holes
(IMBHs), objects whose size falls between that of stellar and SMBHs, in the
range $10^2$--$10^{6} M_{\odot}$. IMBHs are a generic prediction of SMBHs
scenarios, and for which compelling evidence has been recently
obtained~\cite{2014Natur.513...74P}. Mini-spikes around IMBHs are interesting targets
for indirect DM searches, based on the search for secondary particles, like
gamma-rays, neutrinos and anti-matter, produced by the annihilation of WIMPs~\citep{zhao2005, bhs2005, Bertone:2006nq, Fornasa:2007nr, Aharonian:2008wt,
Taoso:2008qz, Bringmann:2009ip, Bertone:2009kj}.  IMBHs may be in particular
hosted at the center of dwarf spheroidal galaxies in the MW halo.  
Dynamical constraints
on their mass are rather loose, due to the paucity of stars in these systems,
but tentative estimates can be obtained by extrapolating astrophysical
relationships between the mass of SMBHs and the properties of the host halos
(e.g.~ref.~\citep{safonova2010}).

We study here the enhancement of the DM annihilation
rate produced by the formation of BHs at the center of dwarf spheroidal galaxies under the assumption that the DM is
made of WIMPs. We show that the lack of excess photons from dwarf spheroidal galaxies in
\Fermi\ LAT satellite data allow to set very stringent limits on mini-spikes in these astrophysical structures,
and estimate the maximum mass of black holes producing them as a function of
WIMP mass (for another recent analysis that instead concentrates on
constraining WIMP DM models see ref.~\cite{gonzalez-morales2014}).

We also perform Monte Carlo simulations of merger histories of black holes in dwarf
spheroidals in a scenario where black holes form the direct collapse of
primordial gas in early halos, and show that this specific formation scenario is 
incompatible at the $84\%$ CL with dark matter being in the form of thermal WIMPs.

The article is organized as follows: in section \ref{sec:minispikes} we discuss
the DM enhancement produced by the formation of a black hole at the center of
dwarf galaxies; in section \ref{sec:simulations} we discuss our simulations of
IMBHs in dwarf galaxies and existing limits; in section \ref{sec:analysis} we
present the \Fermi\ data we used, our analysis technique and its results and we
present our conclusions in section \ref{sec:conclusion}.

\section{Mini-spikes} 
\label{sec:minispikes}

\subsection{Dwarf spheroidals}
\label{sec:dsph}

Dwarf spheroidal galaxies (dSphs) are low luminosity ($L=10^3-10^8 L_\odot$
compared to $L\sim2\times10^{10} L_\odot$ for a normal galaxy) objects orbiting
the MW with masses within their half-light radius on the order of $10^5$--$10^7
M_\odot$~\cite{collins2014}. They are the largest galactic substructures
predicted in the cold DM model and ideal laboratories for DM indirect
detection, for various reasons. They are largely DM dominated systems, as shown
by their high mass-to-light ratios ($M/L\sim100$--$1000\, M_\odot/L_\odot$~\citep{abdo2010}). 
This allows the use of stars as trace-particles of the DM
gravitational potential and thus constraints on the DM distribution can be
derived from stellar kinematics. Moreover, dSphs contain no detected neutral or
ionized gas and show little to no star formation activity~\citep{mateo1998,
gallagher2003, grcevich2009}, which would simplify the interpretation of the
detection of a gamma-ray excess in the direction of a dSph. 

We choose as an illustrative example the Draco dSph, as it has the highest
expected $J$-factor of all the Local Group's dSph satellites (see also
table~\ref{tab:dsph}). Additionally, its relatively high galactic latitude ($b
= 57.9^{\circ}$) makes contamination by other gamma-ray sources in the galactic
disk unlikely. Deep photometric studies of Draco indicate it is unaffected by 
Galactic tides~\cite{segall2007}, which supports our assumption (see next
subsection) of a simple power-law density profile as well as disfavors the
possible disruption of a mini-spike by Galactic tidal forces or merger events.

\begin{table}[!htpb]\centering
    \begin{tabular}{l  c c c c}
        \toprule
        Name & $D$ (kpc) & $r_{s}$ (kpc) & $\rho_{s}$ ($10^8 M_\odot/$kpc$^3$) & $J_\text{NFW}$ $\left(10^{19} \frac{\text{GeV}^2}{\text{cm}^5}\right)$  \\
        \midrule
        \rule{0pt}{3ex}   Bootes I & $62\pm3$ & 0.27 & 2.04 & $0.16^{+0.35}_{-0.13}$ \\
        \rule{0pt}{3ex}   Coma Berenices & $44\pm4$ & 0.16 & 2.57 & $0.16^{+0.22}_{-0.08}$ \\
        \rule{0pt}{3ex}   Draco & $76\pm5$  & 2.09 & 0.26 & $1.20^{+0.31}_{-0.25}$ \\
        \rule{0pt}{3ex}   Fornax & $138\pm8$ & 0.58 & 0.66 & $0.06^{+0.03}_{-0.03}$ \\
        \rule{0pt}{3ex}   Sculptor & $79\pm4$ & 0.95 & 0.37 & $0.24^{+0.06}_{-0.06}$ \\
        \rule{0pt}{3ex}   Sextans & $86\pm4$ & 0.37 & 0.85 & $0.06^{+0.03}_{-0.02}$ \\
        \rule{0pt}{3ex}   Ursa Major II & $30\pm5$ & 0.65 & 0.98 & $0.58^{+0.91}_{-0.35}$ \\
        \rule{0pt}{3ex}   Ursa Minor & $66\pm3$ & 0.17 & 3.47 & $0.64^{+0.25}_{-0.18}$ \\
        \bottomrule
    \end{tabular}
    \caption{Properties of dSphs in the Milky Way halo with the highest expected
    J-factors, from ref.~\cite{abdo2010}.  Here, D is the distance, $r_{s}$ and
    $\rho_{s}$ are respectively the scale radius and density and $J_\text{NFW}$
    is the $J$-factor for a solid angle of $2.4\times10^{-4}$ sr and an NFW profile.}
    \label{tab:dsph}
\end{table}

\subsection{Gamma-rays from DM annihilation}

The density profile of is often approximated with a so-called Navarro, Frenk
and White (NFW) profile~\citep{nfw1997}:
\begin{equation}
    \rho(r) = \frac{r_{s}^3 \rho_{s}}{r (r + r_{s})^2}\;,
    \label{eq:NFW}
\end{equation}
where $r_{s}$ and $\rho_{s}$ are the scale radius and scale density,
respectively, and for Draco we take the values $r_{s} = 2.09\, \text{kpc}$ and
$\rho_{s} = 10^{7.41}\,M_{\odot}/\text{kpc}^3  \cong 0.976\,
\text{GeV}/\text{cm}^3 $~\citep{abdo2010}.  We will discuss below how our results 
depend on this specific assumption. For a self-annihilating DM particle
and fixed gamma-ray energy spectrum, the gamma-ray flux from annihilation
scales with $\rho^2$, specifically $\Phi \propto \frac{\langle\sigma
v\rangle}{m_{\chi}^2} J $, where $m_{\chi}$ is the mass of the WIMP,
$\langle\sigma v\rangle$ is its cross section (assumed to be velocity
independent) and the J-factor is defined as the integral of the DM density
$\rho_{DM}$ squared over the line of sight $l$,
\begin{equation}
    J = \int_{l.o.s.} \rho_{DM}^2 \mathrm{d}l\;.
    \label{eq:J}
\end{equation}
In this case, however, we are interested in the averaged flux over a region 
$\Delta\Omega$ and thus the gamma-ray flux is given by
\begin{equation}
    \frac{\mathrm{d}\Phi}{\mathrm{d}E}(\Delta \Omega) =
    \frac{1}{2}\frac{\mathrm{d}N_\gamma}{\mathrm{d}E} \frac{\langle\sigma
    v\rangle}{m_{\chi}^2} \frac{\bar{J}(\Delta\Omega)}{4 \pi},
    \label{eq:Phi1}
\end{equation}
where $\bar{J}(\Delta\Omega)$ is the average J-factor for a region of
$\Delta\Omega$, and $\mathrm{d}N_\gamma/\mathrm{d}E$ is the photon energy
spectrum.

\subsection{Intermediate mass black holes}
\label{sec:form}

So far, the most convincing observational evidence for the existence of IMBHs
is in the form of the so called ultra-luminous X-ray sources (ULXs). These are
extragalactic, non-nuclear sources of X-ray radiation with isotropic
luminosities in the 0.3--10 keV band of $L_X > 10^{39}\,
\text{erg}\,\text{s}^{-1}$, exceeding those of stellar mass black holes
accreting at the Eddington rate (e.g.~\citep{swartz2004}). Recently, the existence of 
a $400 M_{\odot}$ IMBH powering the X-ray source M82 X-1 has been 
reported in ref.~\citep{2014Natur.513...74P}.
Several theoretical
motivations for the existence of IMBHs have also been raised. 
Strong empirical correlations between the mass of SMBHs and the
properties of their host galaxies suggest an inherent connection between the
growth of the SMBH and the formation and evolution of galaxies. Extending these
relations down to low-mass galaxies predicts the presence of IMBHs in their
centers~\citep{safonova2010}. Additionally, they have been proposed as the initial seeds of SMBHs
(see e.g.~\citep{volonteri2010}).

\medskip

We now briefly review two proposed formation scenarios for IMBHs discussed in
ref.~\citep{bzs2005}. In the first, scenario popIII, black holes form after the
collapse of population III stars. The second scenario pGas features black holes
originating from massive objects formed directly during the collapse of
primordial gas in early-forming halos.

\begin{description}
  \item[Scenario popIII] \hfill \\
  The evolution and eventual fate of massive stars with zero metallicity
  differs significantly from those of their metal-enriched equivalents~\citep{madau2001}. 
  Pop I main sequence stars with masses over 100 $M_{\odot}$
  are vibrationally unstable to radial pulsations, leading to substantial mass
  loss, an effect which is substantially suppressed in low metallicity massive
  stars. This, combined with the fact that, at zero metallicity, mass loss
  through stellar winds is also negligible, means that massive Pop III stars
  could reach the end of their life with a mass close to their initial mass.
  
  However, whereas zero metallicity stars with masses in the range $M_\bullet 
  \sim 60 \text{--} 140 M_\odot$ and $M_\bullet \ge 260 M_\odot$ collapse directly into black holes, 
  those with masses between $140 \leq M/M_{\odot} \leq 260$ will explode in 
  a pair-instability supernova, leaving behind no remnant~\citep{heger2003}.

  \item[Scenario pGas] \hfill \\
  As the first halos virialize and collapse, gas cools and collapses as well,
  forming pressure-supported disks at the center of those halos massive enough
  to contain sufficient amounts of molecular hydrogen. If molecular hydrogen
  cooling is efficient and no major mergers occur over a dynamical time, a
  protogalactic disk can form. Gravitational instabilities in the disk produce
  an effective viscosity that transfer angular momentum outwards and mass
  inwards. By the time this process is halted by the heating of the disk by
  supernovae of Pop III stars, a baryonic mass on the order of $\sim 10^5
  M_{\odot}$ has been transferred to the center of the disk~\cite{koushiappas2004}. 
  Such an object may be temporarily pressure supported, but will inevitably collapse 
  into a black hole~\cite{heger2003, shapiro1983}. 
\end{description}

In both scenarios, the initial black hole forms and grows on timescales long
enough to ensure adiabaticity~\cite{bzs2005}. This causes the surrounding DM
halo to contract, resulting in a mini-spike in the density profile.  However,
the mini-spike is unlikely to survive in the case of off-center (relative to
the DM halo) IMBH formation (i.e. scenario popIII), as we discuss in
section~\ref{sec:persist}.  Thus, we are mostly interested in scenario pGas.

\subsection{DM spiked density profile and annihilation rate}

\begin{figure}
    \centering
    \includegraphics[angle=270, width=0.7\textwidth]{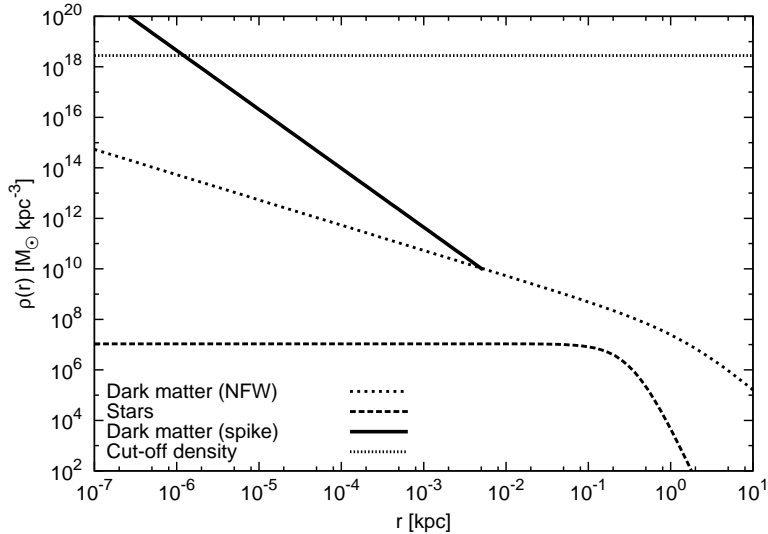} 
    \caption{We show here the different density profiles of Draco as a function
    of radius. For this plot values of $M_\bullet = 10^5\;M_{\odot}$, $t_\bullet=10^{10}$ yr, 
    $m_{\chi}=100\;\text{GeV}$ and $\langle\sigma v\rangle = 3\times10^{-26}
    \;\text{cm}^3 \text{s}^{-1}$ were chosen.}
    \label{fig:rho}
\end{figure}

We will first consider the radius of the gravitational influence of the black
hole $r_{h}$, which is the radius at which the enclosed mass equals twice the
black hole mass~\citep{bzs2005}: 
\begin{equation}
    M(<r_h) \equiv \int_{0}^{r_{h}} \rho_{DM}(r) 4 \pi r^2\mathrm{d}r = 2
    M_\bullet\;,
    \label{eq:rh}
\end{equation}
where $\rho_{DM}$ is the density of DM. 
For an arbitrary power law $\rho(r)=\rho_{s} (r/r_{s})^{-\gamma}$, from
eq.~\eqref{eq:rh} it follows that
\begin{equation}
    r_{h} = r_{s} \left(\frac{3-\gamma}{2 \pi} \frac{M_\bullet}{\rho_s
        r_s^3}\right)^{1/(3-\gamma)}\;.
    \label{eq:rh2}
\end{equation}
For Draco, we can consider both the enclosed mass from the DM in the NFW profile as
well as the enclosed stellar mass, and so eq.~\eqref{eq:rh} becomes
\begin{equation}
    M(<r_h) \equiv \int_{0}^{r_{h}} [\rho_{*}(r)+\rho_{DM}(r)] 4 \pi
    r^2\mathrm{d}r = 2 M_\bullet\;,
    \label{eq:rhdr}
\end{equation}
where $\rho_{*}$ is the density of stellar matter. For Draco we adopt from 
ref.~\citep{mashchenko2006}
\begin{equation}
    \rho_{*}(r) = \rho_{*,0}
    \left(1+\left(\frac{r}{b}\right)^2\right)^{-\alpha/2}\;,
    \label{eq:rhostar}
\end{equation}
with $\alpha = 7$, $b=0.394\, \text{kpc}$ and $\rho_{*,0} = 1.08\times10^7 \,
M_{\odot}/\text{kpc}^3$. Using this, we find for Draco $r_h \sim 0.8\,
\text{pc}$ ($r_h \sim 25\,\text{pc}$) for $M_\bullet = 10^2\, M_\odot$ ($10^5
M_\odot$). For Draco (and dSphs in general) the low stellar density near the
center compared to the DM density (see also figure~\ref{fig:rho}) means that
the difference in $r_h$ calculated with eq.~\eqref{eq:rh} as opposed
to~\ref{eq:rh2} is marginal ($0.6\%$ at most). We will therefor ignore the contribution 
of stellar matter to the radius of gravitational influence when calculating the spiked profile.

As mentioned in section~\ref{sec:form}, we assume the black hole grows
adiabatically. This causes the surrounding DM to contract, leading to an
increase in the DM density. If the BH forms and grows at the center of the DM
halo, then, for an initial density profile like eq.~\eqref{eq:NFW}, the spiked
density profile follows~\cite{gondolosilk1999}:
\begin{equation}
    \rho_\text{sp}(r) =\rho(r_\text{sp})
    \left(\frac{r}{r_\text{sp}}\right)^{-\gamma_\text{sp}}\;,
    \label{eq:rhosp}
\end{equation}
where the radius of the spike is $r_{sp} \approx 0.2 r_{h}$~\citep{merritt2004}
and $\gamma_\text{sp} = (9-2\gamma)/(4-\gamma)$. We therefore see that $\gamma_\text{sp}$
is a weak function of $\gamma$ and  for an initial NFW (with $\gamma=1$)
$r_\text{sp} \approx 5\;\text{pc}$ and, $\gamma_\text{sp}= 7/3$. The result can be easily generalised 
to other DM profiles.  

Although formally divergent, a physical upper limit to the
DM halo density arises for self-annihilating WIMP models:
\begin{equation}
    \rho_\text{lim} \equiv \frac{m_{\chi}}{\langle\sigma v\rangle \;
    (t-t_{f})} \;,
    \label{eq:rlim}
\end{equation}
where $t-t_{f}=t_\bullet$ is the age of the halo, which for Draco we estimated
to be $10^{10}$~yr, based on the chemical composition of its stellar population~\cite{castellani1975}. 
This shows that annihilations place an upper limit on the
DM density of the order $m_\chi/\langle\sigma v\rangle(t-t_f)$, and we define
$r_\text{cut}$ as the radius where $\rho_{sp}(r_\text{cut}) =  \rho_\text{lim}$.\footnote{
Strictly speaking, the cut-off only appears in the case of circular DM particle 
orbits, see~\citep{2007PhRvD..76j3532V}, in other cases a weak cusp is formed. 
The difference is negligible for our case.}
For $m_{\chi} = 100\;\text{GeV}$ and $\langle\sigma v\rangle = 3\times10^{-26}
\text{cm}^3\text{s}^{-1}$ we find $\rho_\text{lim} = 1.1\times10^{11}
\text{GeV}/\text{cm}^3 \cong 2.8 \times 10^{18} M_{\odot}/\text{kpc}^3 $ (see
also figure~\ref{fig:rho}) and a black hole mass of $10^5 M_{\odot}$ then gives
$r_\text{cut} = 6.3\times10^{-4} \text{pc}$.

Finally, we can express the flux of gamma-rays from a spike around an IMBH as
\citep{bzs2005}:
\begin{equation}\label{eq:Phi}
    \frac{\mathrm{d}\Phi_\text{spike}}{\mathrm{d}E} \simeq \frac{1}{2}
    \frac{\langle\sigma v\rangle}{m_{\chi}^2} \frac{1}{D^2}
    \frac{\mathrm{d}N_\gamma}{\mathrm{d}E} \int_{r_\text{cut}}^{r_{sp}}
    \rho_\text{sp}^2(r) r^2 \mathrm{d}r \simeq
    \frac{10}{3}\frac{\mathrm{d}N_\gamma}{\mathrm{d}E} \frac{\langle\sigma
    v\rangle}{m_\chi^2} \frac{\rho^2(r_\text{sp})}{D^2} r_\text{sp}^{14/3}
    r_\text{cut}^{-5/3},
\end{equation}
where in the last step we assumed $r_\text{sp} \gg r_\text{cut}$. The total
flux originating from a spiked profile is the sum of the flux of the mini-spike
and that of the NFW profile. 
%

\bigskip

To extract the functional dependence of $\Phi_\text{sp}$ on $M_\bullet$, we
approximate $r_h$ by using eq.~\eqref{eq:rh2} with $\gamma=1$. In doing so, we
neglect the contribution of stars and the cusp of the NFW profile. However,
these contribute little to the density at a radius on the order of $r_h \sim
10^{-3} \text{kpc}$ (see also figure~\ref{fig:rho}). Using $r_\text{sp}=0.2 r_h
= 0.2 \sqrt{M_\bullet/\pi \rho_0 r_0}$ and
$r_\text{cut}=\left(r^4_\text{sp}\left(\rho_0 r_0
t_\bullet/m_\chi\right)^3\right)^{1/7}$ and substituting these into
eq.~\eqref{eq:Phi}, we find
\begin{equation}\label{eq:phiscale}
    \Phi_\text{spike} \propto \frac{\langle\sigma v\rangle^{2/7}}{m^{9/7}_\chi}
    \frac{M^{6/7}_\bullet r^{3/7}_0 \rho^{3/7}_0}{t_\bullet^{5/7}}.
\end{equation}
From this, we see that, for a fixed $\Phi_\text{spike}$, $\langle\sigma
v\rangle \propto M^{-3}_\bullet$. 

\subsection{Persistence of the mini-spike}
\label{sec:persist}

There are several possible events in the history of a dSph that could have the
effect of lessening or completely erasing the density spike or preventing it
from forming at all. For instance, simulations have shown that a merger of two
DM halos, each containing a MBH, leads to a BH binary which ejects DM particles
that pass the binary within a few times its semi-major axis \cite{merritt2002}.
In certain cases, this might even have the effect of decreasing the DM density
below that of a halo without an IMBH. Other phenomena, for example the passing
of a molecular cloud, open/globular cluster or the presence of a central bar
would all serve to heat up the cold orbits of the DM particles in the spike,
diminishing or destroying the spike \cite{uzk2001}.

Additionally, ref.~\cite{uzk2001} examined the importance of the assumptions
that the BH grows adiabatically. In the case that the BH growth occurs too
rapidly to be considered adiabatic (approximated by instantaneous growth of the
BH), the angular momentum of an orbiting DM particle is still conserved, but
its energy is not. Unlike the case of adiabatic growth, this means that an
initially circular orbit does not stay circular, instead DM particles will
follow elliptical, Keplerian orbits. While particles in these orbits reach
smaller radii than the circular orbits in the adiabatic case, their velocities
at these small radii are much larger, the result of which being that a DM
particle spends most of its time at radii larger than in the adiabatic case. It
follows that, while the BH still causes a density enhancement, it is much
shallower than in the adiabatic case: $\rho_\text{sp}\propto r^{-4/3}$ for an
initial NFW profile. Thus, for a non-adiabatically growing IMBH, eq.~\eqref{eq:Phi}
becomes
\begin{equation}
    \Phi_\text{spike, non-ad.} =
    \frac{3}{2}\frac{\mathrm{d}N_\gamma}{\mathrm{d}E} \frac{\langle\sigma
    v\rangle}{m_\chi^2} \frac{\rho^2(r_\text{sp})}{D^2} r_\text{sp}^{3},
\end{equation}
where we again use $r_\text{cut} \ll r_\text{sp}$.

In the case of an off-center (with respect to the DM halo) formation of an
IMBH, like in case of the popIII scenario, the spike is also significantly shallower, with $\rho \propto r^{-3/2}$ \cite{gondolosilk1999, zhao2005}, and the resulting flux is given by
\begin{equation}
    \Phi_\text{spike, off-c.} =
    \frac{1}{2}\frac{\mathrm{d}N_\gamma}{\mathrm{d}E} \frac{\langle\sigma
    v\rangle}{m_\chi^2} \frac{\rho^2(r_\text{sp})}{D^2}
    r_\text{sp}^{3}\;\text{log}\!\left[\frac{r_\text{sp}}{r_\text{cut}}\right].
\end{equation}

\section{Intermediate mass black holes in dwarf spheroidals}
\label{sec:simulations}

Previous analyses of mini-spikes have focused on randomly generated populations
of IMBHs, in the framework of specific formation scenarios, and on comparisons
with gamma-ray \cite{bzs2005,Fornasa:2007ap,Bringmann:2009ip,bertone2009} and
neutrino \citep{Bertone:2006nq} observations. Here we consider specifically
dwarf galaxies, for which halo properties are observationally know, and assume
the mass of the black hole as a free parameter. 

\subsection{Monte Carlo simulation of merger histories}
\label{sec:montecarlo}

Refs.~\cite{uzk2001, merritt2002} studied DM spikes as a result of the presence
of a SMBH in the context of a MW-like galaxy, and discuss the many incidents
that would disrupt a spike at the Galactic center. As noted above events that
are connected to the evolution of a galaxy in hierarchical structure formation,
such as BH binary mergers and rapid gas accretion on the BH would decrease the
relevance of a DM spike.\footnote{For a recent discussion of a spike at the 
Galactic center in the context of the 1--3 GeV excess, see \citep{2014arXiv1406.4856F}} 
To estimate the impact of BH mergers and gas accretion
we develop models of the hierarchical evolution of dSph galaxies inside a
MW-like galaxy, and trace their dynamical and accretion history. The models are
presented in detail in ref.~\cite{2010MNRAS.408.1139V} and we summarize here
their main features. 

\medskip

Our technique follows that of \cite{VHM}, as we use Monte Carlo realizations of
the merger histories of DM halos. We analyze here 15 different realizations  of
halos that reach a mass of $M_h=2 \times10^{12}$ M$_\odot$ at $z=0$. We seed
the  high-redshift progenitor halos with BHs as per scenario popIII and
scenario pGas and follow them from formation to $z=0$. 

For scenario popIII we assume that Pop III stars form in metal-free halos with
$T_{\rm vir}> 2000$ K \citep{Yoshida2006}, where we model metal enrichment by
the `high feedback, best guess' model of ref.~\cite{Scannapieco2003}.  Namely,
we model metal enrichment via pair-instability supernovae winds, by following
the expansion of spherical outflows into the Hubble flow and calculate the
probability that a halo of mass $M_h$ forms from metal-free gas at a redshift
$z$. When a halo forms in our merger tree we calculate the probability that it
is metal-free. If a random draw satisfies this conditions we assume a seed BH
of 100\,M$_\odot$ forms.  

For scenario pGas we rely instead on the collapse of of dense gas, where we
assume that gas is accumulated in the center of a halo via viscous
instabilities \citep{LN2006,VLN2008}. The gas inflow can be computed from the
Toomre stability criterion (via the critical Toomre parameter, $Q_{\rm
c}\simeq2$).  This process is particularly effective for metal--free halos
where cooling is driven by atomic hydrogen cooling (we test for metal
enrichment of a halo as in scenario popIII). Atomic hydrogen cooling is
effective in systems with $T_{\rm vir} > 10^4$K, corresponding to large halos
with mass $\approx  10^8$ M$_\odot$. The gas can cool via atomic line cooling
to a temperature $T_{\rm gas}=5000$\,K.  The efficiency of the seed assembly
process ceases at large halo masses ($T_{\rm vir}>1.4\times10^4$ K), where the
mass-accretion rate from the halo is above the critical threshold for
fragmentation and the disc undergoes global star formation instead.  Given a DM
halo with halo mass $M_{\rm h}$, virial
temperature $T_{\rm vir}$, and spin parameter $\lambda$ \citep{MoMaoWhite1998},
the BH mass can be expressed as \citep{LN2006}: 
\begin{equation}
    M_\bullet= f_{\rm d}M_{\rm h}\left[1-\sqrt{\frac{8\lambda}{f_{\rm d}Q_{\rm
    c}}\left(\frac{j_{\rm d}}{f_{\rm d}}\right)\left(\frac{T_{\rm gas}}{T_{\rm
    vir}}\right)^{1/2}}\right], 
\end{equation}
for $\lambda<\lambda_{\rm max}=(f_{\rm d}Q_{\rm c}/8)(f_{\rm d}/j_{\rm d})
(T_{\rm  vir}/T_{\rm gas})^{1/2}$. Here $\lambda_{\rm max}$ is the maximum halo
spin parameter for which the disc is gravitationally unstable,  $f_{\rm d}\sim
0.05$ normalizes the amount of gas that participates in the infall expressed as a fraction of the halo mass, and $j_{\rm d}\sim
f_{\rm d}$ is the fraction  of the halo angular momentum retained by the
collapsing gas.   

Throughout the BH+halo evolution we keep track of BH-BH mergers prompted by
halo mergers, and merger-driven gas inflows that can trigger rapid growth of
the BH and disrupt the DM spike.  We define the redshift of the latest such
accretion episode as the ``age" of the BH. 

\medskip

Finally, from the population of MW satellites that survive to $z=0$ we have to
extract those that resemble the dSphs analyzed in this paper. We define a given
satellite in our simulation as dSph analog if it matches a given dSph in DM
halo mass and distance from the Galactic center. This requires in the first
place to model the dynamical evolution of a satellite. When a satellite enters
the halo of the MW it evolves in the potential well of the host until $z=0$,
experiencing tidal stripping and possibly merging with the host.  For each
satellite that merges with the main halo of the merger tree, we evolve the
satellite-host system by integrating the equation of motion of the satellite in
the gravitational potential of the host (assuming spherical NFW profiles,
\cite{nfw1997}).  In the equations of motion, we include the dynamical friction
term:
\begin{equation}
    {d^2 {\vec r} \over d t^2} =  -{G M(r) \over r^2}\, {\vec r} - 
    {4 \pi G^2 \ln\Lambda\,\rho\, M_{\rm sat} \over v^2} f(x)\,{\vec v},
    \label{DF}
\end{equation}
where $f(x)\equiv [{\rm erf}(x) - (2 x/\sqrt{\pi}) e^{-x^2}]$, $x\equiv
v/\sqrt{2}\sigma$, and the velocity dispersion $\sigma$ is derived from the
Jeans' equation for the composite density profile, assuming isotropy
\cite{Binney1987}. Here $M(r)$ describes the total  mass of the host within
$r$, $\rho(r)$ is the total density profile, and the second term represents
dynamical friction against the background. We include the BH Keplerian
potential if the galaxies host a BH.  The Coulomb logarithm, $\ln \Lambda$, in
eq.~\eqref{DF} is taken equal to 2.5  \citep{Taylor2001}. The mass of the
satellite evolves during the integration because of tidal stripping. At every
step of the integration we compare the mean density of the satellite to the
mean density of the host halo at the location of the satellite. Tidal stripping
occurs at the radius within which the mean density of the satellite exceeds the
density of the galaxy interior to its orbital radius \citep{Taylor2001}.   We
trace the orbital evolution and the tidal stripping of all satellites from the
time when the satellite enters the virial radius of the host to $z=0$.
Satellites that survive until the present time provide an analogue of the dSph
population around the MW (see ref.~\cite{2010MNRAS.408.1139V} for a comparison
with observational properties of the dSph population).

\subsubsection{IMBHs in dwarf spheroidals of the Milky Way and M31}
\label{sec:MWM31}

Given our assumptions for BH formation, not all galaxies necessarily  host a
BH. In the case of Draco-analogs we find \emph{that only 18\% of them host a
central BH in scenario popIII, and 8\% in scenario pGas}.  However, when we
analyze the population of BHs in simulated dSph we find that many of the
incidents studied by \cite{uzk2001} and \cite{merritt2002} that could diminish the spike in the case of
the Milky-Way, e.g. DM scattering off stars, do not apply. Our simulations of Draco-like halos show $57\%$ of
BHs in scenario popIII and $35\%$ in scenario pGas undergo BH-BH mergers, with
the mean redshift since the last merger being $z=7.2\pm4.0$ for scenario popIII
and $4.3\pm3.2$ for scenario pGas, at $t\sim0.8\;\text{Gyr}$ and
$t\sim1.4\;\text{Gyr}$, respectively.  

For comparison, the fraction of Ursa Minor-analogs that host a central BH are
similar to Draco's, while in the case of Ursa Major II the fractions increase
to 41\% and 14\% in scenario popIII and pGas respectively. The fraction of BHs
that experienced a BH-BH merger remains between 30\% and 60\%. The BH ages are
also comparable to Draco's case. 

\medskip

We can use our simulations to derive an estimation on the probability of a dSph
in the MW or the M31 (Andromeda) galaxy containing a scenario popIII/pGas IMBH.
This is done by taking a \emph{population of 30 dSphs} similar to that of the MW (18 dSphs, as taken from
\cite{fermi2013}) and M31 (12 dSphs, from \cite{splash2012}) from our simulations. 
We consider a simulated dSph as an analog to a physical dSph if the distance to
the center of its host galaxy lies within 10\% of its physical counter-part.
We assume all dSphs have a typical halo mass of $\sim 3\times10^9 M_\odot$~\citep{wolf2010}, 
and look for analogs with masses within a range of $3\times10^8M_{\odot} \le M_h \le 3\times10^{10} M_{\odot}$. 
The probability of forming an IMBH via a scenario is derived by comparing the amount 
of simulated analogs with an IMBH to the total number of analogs. 

Considering the full set of MW and M31 dwarfs, we find in the case of scenario
pGas that the \emph{average} probability of a dSph hosting an IMBH is 10\%, of
which 35\%, on average, undergo a BH-BH merger.  After calculating both the
total probability of a BH being formed and the
total probability of it surviving without a merger event, our simulations
show that the probability of none of the dSphs in either the MW or Andromeda
galaxy hosting an IMBH is 0.053\% (3.4\%) for scenario popIII (pGas),
respectively (see table~\ref{tab:probs}).  The probability of \emph{none} of
the dSphs containing an \emph{unmerged} BH (for which we expect an increased
annihilation flux due to the mini-spike generation as discussed above) is
somewhat higher and 8.5\% (16\%) in scenario popIII (pGas). We checked that both increasing and 
decreasing the assumed halo mass results in an even slightly lower probability that none of the 
dSphs in our simulations host an unmerged pGas IMBH. 

\begin{table}[!htpb]\centering
    \begin{tabular}{l c c}
        \toprule
        & Scenario popIII & Scenario pGas  \\
        \midrule
        No IMBH in any dSph & $5.3\times10^{-4}$ & 0.034 \\
        No unmerged IMBH in any dSph & 0.085 & 0.16  \\
        \bottomrule
    \end{tabular}
    \caption{Total probabilities for the MW and M31 dSphs to \emph{not}
    contain any (unmerged) IMBH, for both IMBH formation scenarios, as
    resulting from our simulations.}
    \label{tab:probs}
\end{table}

\subsection{Other limits on the black hole mass}
\label{sec:otherlims}

We will now examine two alternative methods used to derive estimates of
$M_\bullet$; the $M_\bullet$--$\sigma$ relation and substructure
considerations. We will compare these values to upper limits derived from
gamma-rays in the next section.

\subsubsection{SMBH-host galaxy relations and substructure considerations}
\label{sec:msigma}

There is an empirical relation between the mass of a SMBH and the velocity
dispersion of its host galaxy. This $M_\bullet$--$\sigma_*$ relation is obtained
by fitting a sample of dynamically measured black hole masses and corresponding
line-of-sight velocity dispersions to the power law
\begin{equation}
    \text{log}_{10} M_\bullet = \alpha +\beta\;\text{log}_{10}\left[
    \frac{\sigma_*}{100\;\text{km}\;\text{s}^{-1}}\right],
\end{equation}
where $\alpha=6.91$ and $\beta=4$~\citep{tremaine2002}. Thus, an average velocity 
dispersion of the stellar component for Draco of $9.1$ km/s implies a $M_\bullet$ 
of $557 M_\odot$. It should be noted that the behavior of the $M_\bullet$--$\sigma_*$ 
relation at low $M_\bullet$, or whether it applies at all, is unknown.

In the case of Ursa Minor, an upper limit on $M_\bullet$ has been derived
through N-body simulations. The dwarf UMi features a dynamically cold,
long-lived substructure on the northeast side of its major axis. In the case
without a central IMBH, this substructure can persist long enough to explain
its observation. However, the introduction of a central IMBH causes the clump
to diffuse through gravitational interaction with the BH at a rate depending on
the mass of the BH. If the clump is a primordial feature of UMi, its
persistence imposes an upper limit of $M_\bullet = (2 - 3)\times10^4 M_\odot$
on a central IMBH \cite{lora2009}.

\subsubsection{Kinematic measurements}

We have developed kinematic models of Draco to test the ability to detect (and
possibly measure) a putative BH mass through kinematic measurements. The basic
idea (see refs.~\cite{Ferrarese2005,2013ARA&A..51..511K} for details) is that a
massive BH in a galaxy center causes gas and stars in its vicinity to move
faster than they would in the presence of the galaxy potential only, as an
additional Keplerian potential exists. For all galaxies other than the MW we
have to rely on integrated stellar dynamics. The whole potential of the galaxy
is modeled to extract the signature of the BH, from spectra of the central
nucleus that provide information on the velocity dispersion integrated along
the line of sight vs radius, and the stellar surface brightness. This data is
used to reconstruct the underlying gravitational potential that produces a
self-consistent description of the stellar system, and derive the mass density
(luminous + BH + DM) which reproduces the observables. One such attempt was
done on one dSph, Fornax \citep{2012ApJ...746...89J}, and the 1-sigma upper
limit on the mass of a putative BH in Fornax is $3.2\times 10^4$ M$_\odot$.

We create a simple framework that allows us to estimate the impact of a BH on
the line-of-sight (LOS) velocity dispersion of Draco, and compare it to the
observations \citep{2007ApJ...667L..53W}. We follow a procedure similar to the
one described by \cite{mashchenko2006} except that we treat the whole problem
analytically, by solving Jeans' equation
\citep{Binney1987,1982MNRAS.200..361B,2003MNRAS.343..401L}: 
\begin{equation}
    \frac{d(\rho_* \sigma_r^2)}{dr} + \frac{2\beta}{r} \rho_*= -\rho_* \frac{d\Phi}{dr},
    \label{jeanseq}
\end{equation}
where $\beta(r)\equiv 1-\sigma_t^2(r)/\sigma_r^2(r)$ is the anisotropy
parameter ($\sigma_t$ and $\sigma_r$ are the tangential and radial velocity
dispersions respectively;  $\beta$ equals $-\infty$, 0, and 1 for purely
tangential, isotropic, and purely radial orbits respectively), and $\Phi$ is
the gravitational potential.

Eq.~\ref{jeanseq} admits a solution of the following form:
\begin{equation}
    \rho_* \sigma_r^2(r)=\int_r^\infty \rho_*(s) \frac{G\,M(<s)}{s^2}
    \exp\left[2\int^{s}_r\frac{\beta(t)dt}{t}\right] dt.
\end{equation}
Here $M(<r)$ is the mass within the radius of interest, and includes DM, stars,
and cases without and with a central $10^5$ M$_\odot$ BH.  This is a very large
BH for a galaxy like Draco where the total stellar mass is $\sim 10^6$
M$_\odot$ (normally the BH mass is few $\sim 10^{-3}$ the bulge mass, therefore
$\leqslant$ few $\sim 10^{-3}$ the total stellar mass), but we want to look at
the kinematic signature of an extreme case. A lower mass BH's signature would
only be smaller.

We fix the DM density profile to the one given in Table~\ref{tab:dsph}, and we
adopt a stellar density profile with the functional form given in section 2.4,
but we allow the parameters, $\rho_{*,0}$ and $b$ to vary. We consider both
models with $\beta$ constant with radius (as done by ref.~\cite{fermi2010}),
and models with radius-dependent $\beta$, adopting the functional form
suggested by \cite{mashchenko2006}: $\beta=2\eta/(1+\eta)$, where $\eta =
\eta_0 + (\eta_1-\eta_0)[1-\left(\rho_*/\rho_0)^{1/\lambda}\right]$, and
$\eta_0$, $\eta_1$ and $\lambda$ are free parameters.

We then calculate the LOS velocity dispersion:
\begin{equation}
    \sigma_{LOS}(R)=\frac{2}{\Sigma(R)}\int_R^\infty\left(1-\beta\frac{R^2}{r^2}\right)\frac{\rho_*
    \,\sigma_r^2\, r}{\sqrt{r^2-R^2}}dr,
\end{equation}
with $\Sigma$ being the stellar surface density. 
\begin{figure}[!thbp]\centering
  \includegraphics[angle=0,width=0.6\textwidth]{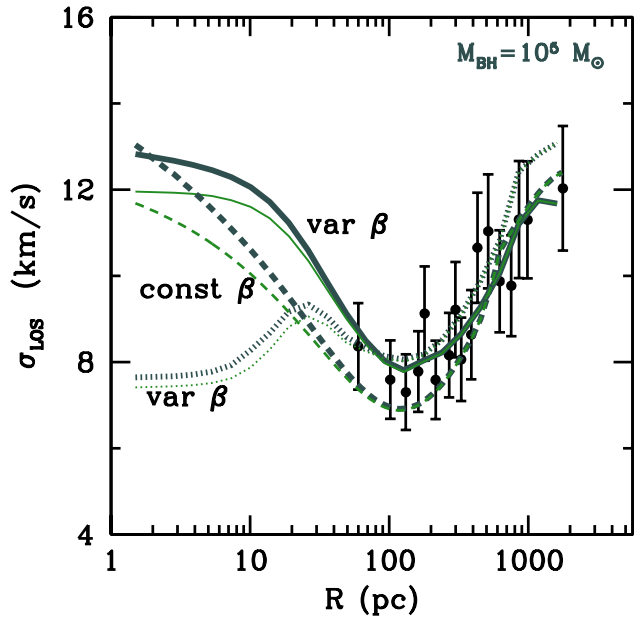}\\
  \caption{{Line of sight velocity dispersion for Draco without (green thin
  curves) and with (dark gray thick curves) a central $10^5$ M$_\odot$ BH. The
  dashed curves refer to a model with anisotropy parameter, $\beta$, constant
  with radius ($\beta=0.3$; $\rho_{*,0}=10^7$ M$_\odot$ , $b=0.125$ kpc), the
  solid and dotted curves to models where $\beta$ is a function of radius
  ($\eta_0=0$; $\eta_1=0.2$; $\lambda=1.7$; $\rho_{*,0}=2.5\times10^7$
  M$_\odot$, $b=0.158$ kpc and $\eta_0=-0.8$; $\eta_1=0$; $\lambda=1.3$;
  $\rho_{*,0}=2\times10^7$ M$_\odot$, $b=0.158$ respectively; for details see
  the text and ref.~\cite{mashchenko2006}).}}
  \label{fig:kin}
\end{figure}
It is well known that the velocity anisotropy, $\beta(r)$, is degenerate with
the mass profile in determining the velocity dispersion, and that projection
effects in determining the LOS velocity dispersion affect the final result. In
fact, we find a large family of models with different  $\rho_{*,0}$, $b$ and
$\beta(r)$ to match the data equivalently well. Additionally, as shown in
fig.~\ref{fig:kin} the signature of the BH is completely invisible in today's
data.  Even if future measurements were to further extend to the inner radii,
the difference between a case with and without BHs is sufficiently small that
extracting the BH mass from the total potential may be daunting, especially
because dSph galaxies have very low central stellar density, therefore very few
stars that one can use as kinematic tracers \citep{2010MNRAS.408.1139V}.

\section{\Fermi\ LAT analysis}
\label{sec:analysis}

We will now update the upper limit derived on the gamma-ray flux from Draco that 
were previously presented in the literature (see e.g.~ref.~\cite{fermi2013}), 
and discuss the impact on and the relevance for the IMBHs in scenario pGas.

\subsection{Data reduction and statistical method}

For our analysis we used data from the \Fermi\ LAT gamma-ray space telescope.
The data used was obtained between 04-08-2008 and 23-06-2014, spans an energy
range of 100 MeV to 100 GeV and it encompasses a $10^{\circ}$ radius circle
centered on the coordinates of Draco; $l = 260.1^{\circ}$, $b = 57.9^{\circ}$.
Using the \Fermi\ Science Tools, we first applied several cuts to the data. We
set the maximum zenith angle to $100^{\circ}$ to exclude time intervals with
elevated background levels caused by intersection of the region of interest
with the Earth's limb, and our cut on rocking angle $< 52^{\circ}$ serves the
same purpose. We also excluded periods when a spacecraft event could have
affected the data quality or the LAT instrument was not in normal science
data-taking mode.

We used the \Fermi\ Science Tools\footnote{Version v9r33p0, freely available
from \url{http://fermi.gsfc.nasa.gov/ssc/data/analysis/software/}} to perform a
binned likelihood analysis, fitting the data to our different models, using
twenty different WIMP masses logarithmically spaced between 6 GeV and 1000 GeV,
for annihilation channels $\overline{b}b$, $\overline{t}t$, $W^{+}W^{-}$,
$e^{+}e^{-}$ and $\mu^{+}\mu^{-}$. For the energy spectrum $dN_\gamma/dE$ for
different masses and annihilation channels, we used tables provided by
ref.~\cite{pppc}. We model Draco as a point source, with a spectrum defined by
the aforementioned $dN_\gamma/dE$ distribution multiplied by a normalization
prefactor. Our analysis resulted in a 95\% confidence level (i.e. $2\Delta
\text{ln}\left(\mathcal{L}\right)=2.71$ where $\mathcal{L}$ is the likelihood) upper limit on the
gamma-ray flux from Draco.

\subsection{Results for Draco}
\label{results}

\subsubsection{Flux and cross section upper limits}
\label{sec:fluxlims}

\begin{figure}\caption{}
    \begin{subfigure}[!hbt]{0.5\textwidth}
        \includegraphics[angle=270,width=\textwidth]{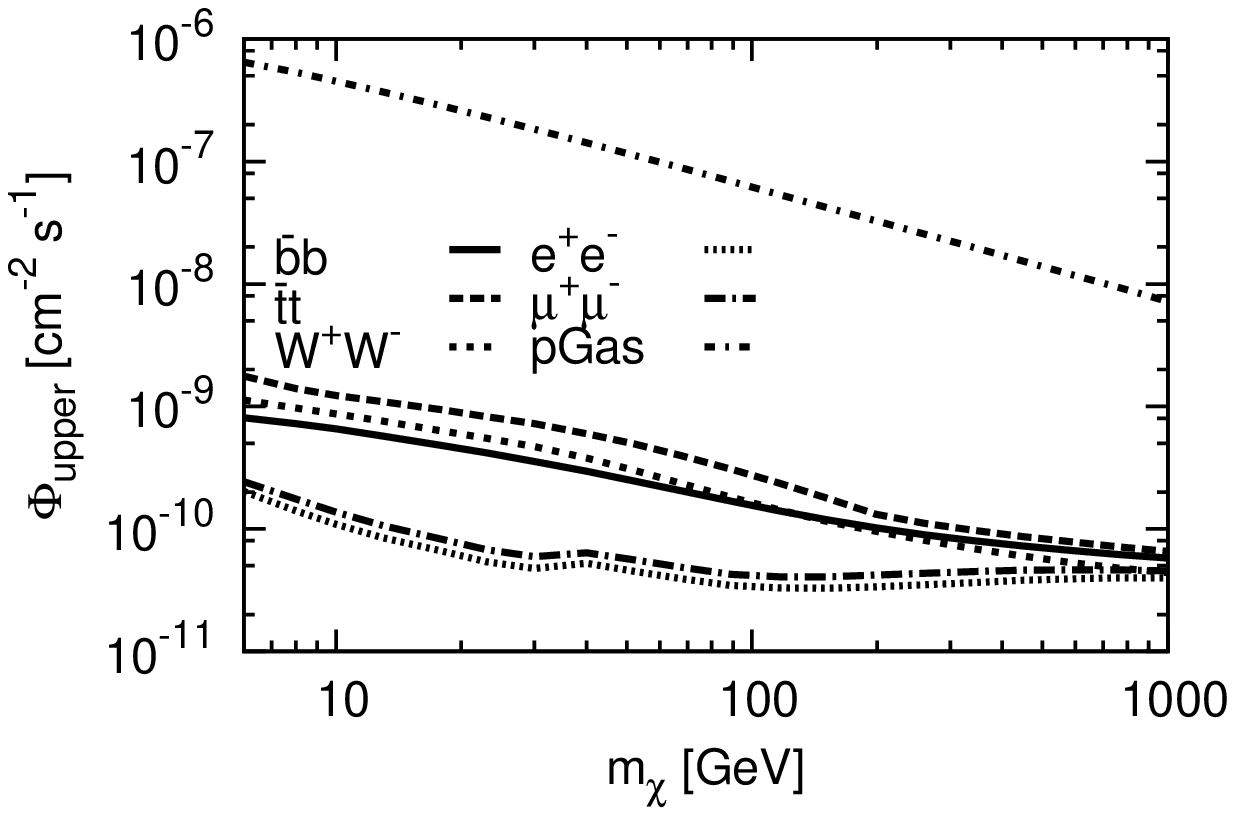}\\
        \caption{Upper limits on gamma-ray flux from WIMP annihilation in
        Draco, for several annihilation channels. For comparison, the expected flux 
        from annihilation into $\overline{b}b$ for a pGas IMBH ($M_\bullet = 10^5 M_\odot$) is also shown.}
        \label{fig:lims}
    \end{subfigure}\;
    \begin{subfigure}[!hbt]{0.5\textwidth}
        \includegraphics[angle=270,width=\textwidth]{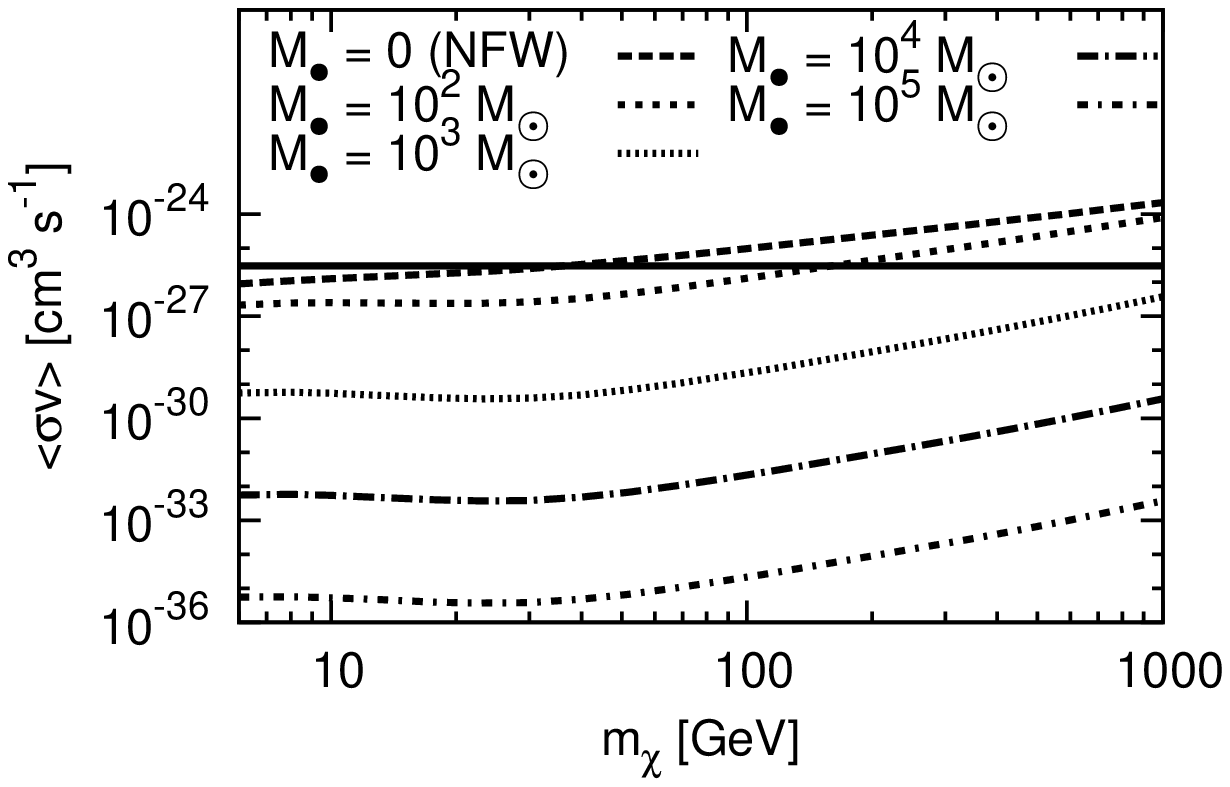}\\
        \caption{Upper limits on WIMP parameters obtained through our analysis
        for annihilation into $\overline{b}b$. The solid line indicates
        $\langle\sigma v\rangle=3\times10^{-26} \text{cm}^3 \text{s}^{-1}$. \\ \\}
        \label{fig:msigma}
    \end{subfigure}
\end{figure}

Our results for the flux upper limits are shown in figure~\ref{fig:lims}, and
are consistent with those previously obtained for example in
ref.~\cite{fermi2013}.  In the case of a typical scenario pGas IMBH with
$M_\bullet \sim 10^5 M_\odot$, the expected flux is between 2 and 3 orders of
magnitude higher than these upper limits, and thus, in the case of a standard
thermal WIMP, scenario pGas IMBHs can be ruled out for all Galactic dSphs.  We
demonstrate this by using our derived flux upper limits to put upper limits on 
the cross section $\langle\sigma v\rangle$ as function of the WIMP mass $m_{\chi}$ and for a certain black hole
mass $M_\bullet$. The results for the $\overline{b}b$ channel and five
different black hole masses is shown in figure~\ref{fig:msigma}. 

\subsubsection{Black hole mass upper limit}


\begin{figure}[!tbp]
    \centering
    \caption{A comparison of the upper limits on black hole mass $M_\bullet$ in
    Draco as a function of WIMP mass $m_\chi$ for the different cases we
    considered.}
    \label{fig:bhcomps}
    \begin{subfigure}[!htbp]{0.45\textwidth}
        \includegraphics[angle=270, width=\textwidth]{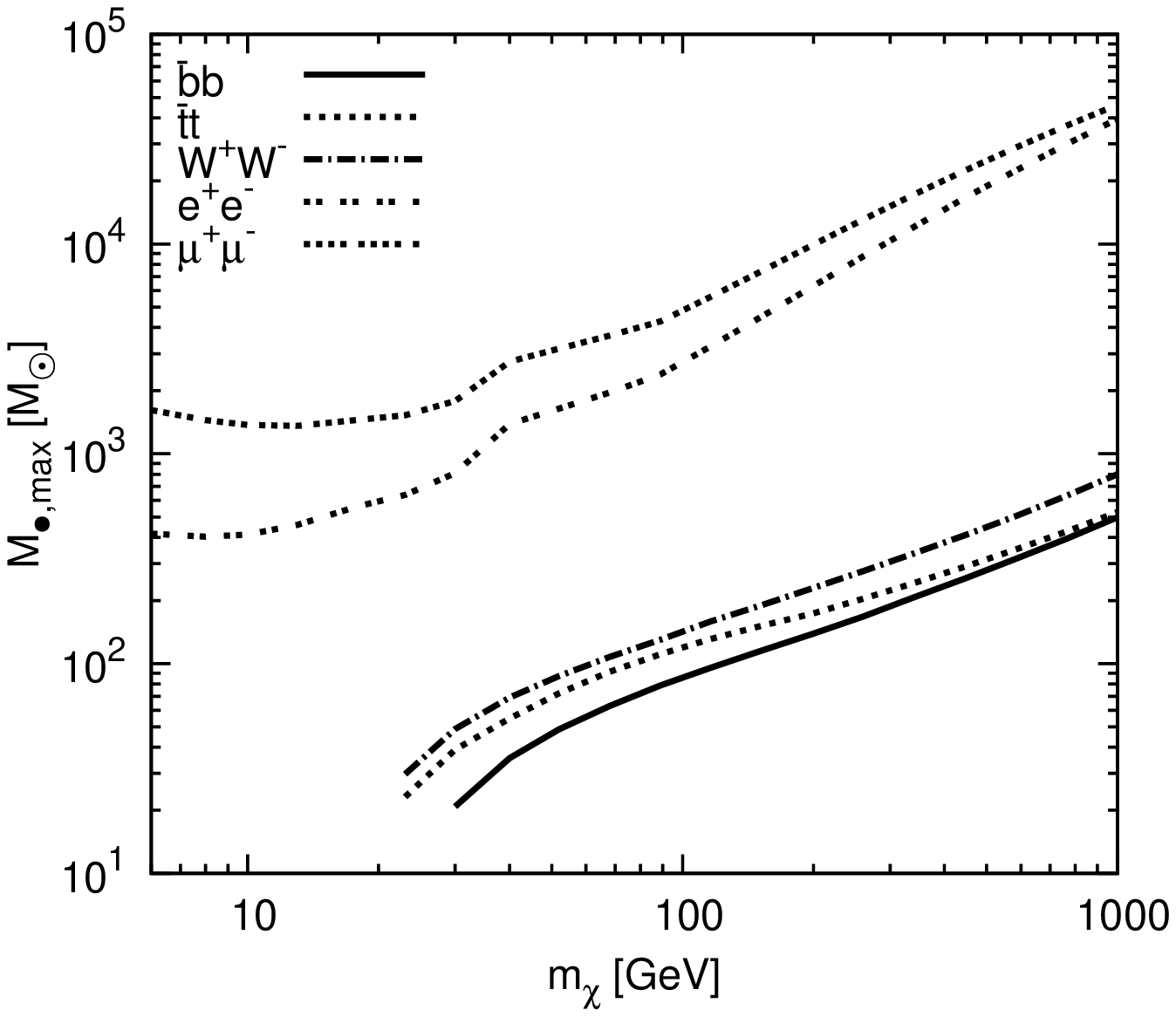}
        \caption{Draco, $\langle\sigma v\rangle=3\times10^{-26} \text{cm}^3
        \text{s}^{-1}$.}
        \label{fig:bhcomp}
    \end{subfigure}\;
    \begin{subfigure}[!htbp]{0.45\textwidth}
        \includegraphics[angle=270,  width=\textwidth]{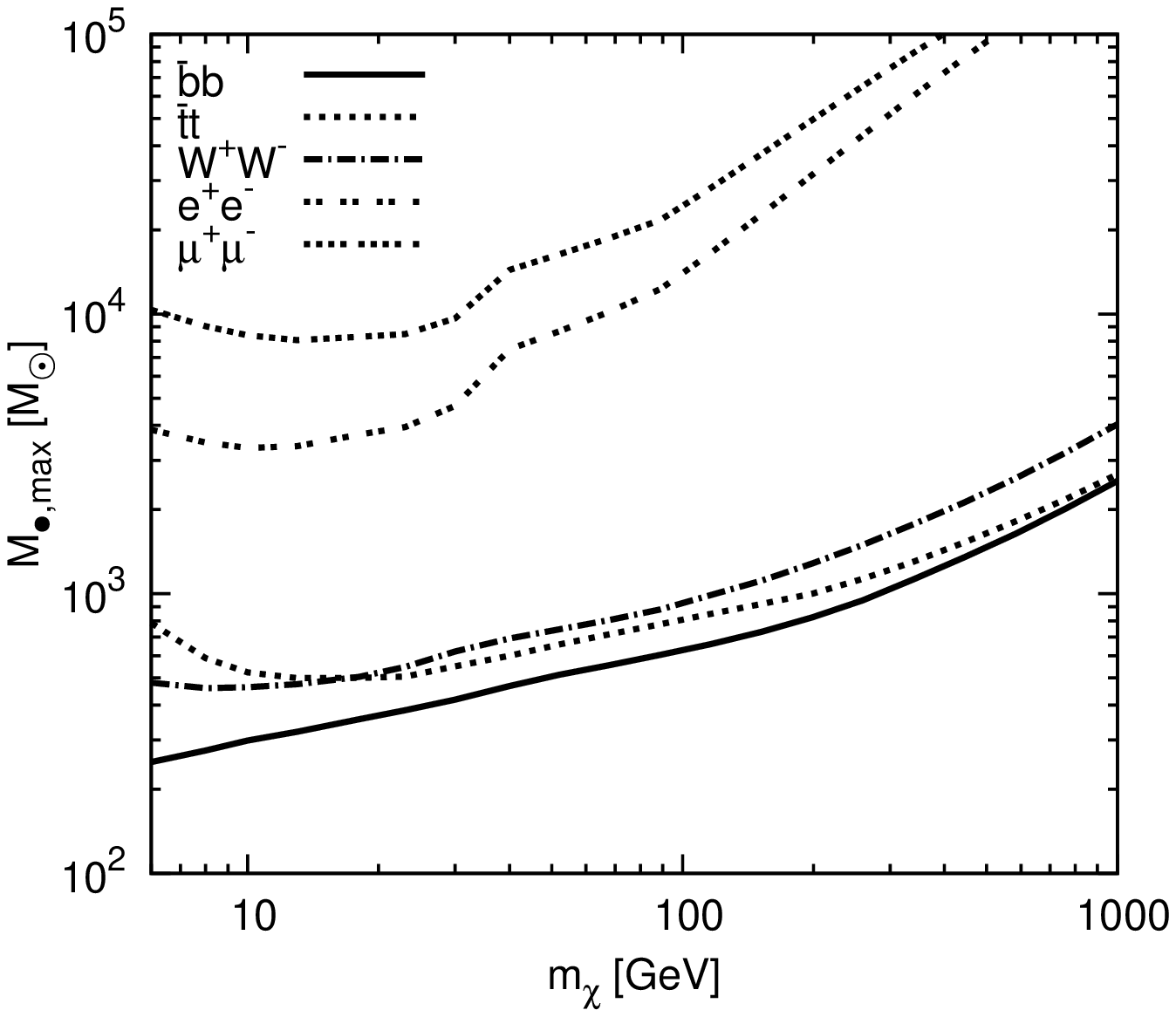}
        \caption{Draco, $\langle\sigma v\rangle=10^{-29} \text{cm}^3
        \text{s}^{-1}$.}
        \label{fig:bhcompalt}
    \end{subfigure}
\end{figure}

Since the annihilation flux increases in general with the mass of the IMBH, we
can derive, for a fixed annihilation cross section and DM mass, upper limits on
the IMBH mass.  To this end, we shall consider two values for the cross
section; the thermal cross section, derived from the assumption that WIMP DM is
a thermal relic from the early universe, $\langle\sigma
v\rangle=3\times10^{-26} \text{cm}^3 \text{s}^{-1}$, and a more ``pessimistic''
scenario (so called because a lower cross section diminishes the prospects for
indirect detection), with $\langle\sigma v\rangle=1\times10^{-29} \text{cm}^3
\text{s}^{-1}$.  As shown in fig.~\ref{fig:bhcomp}, for both cross sections we
rule out IMBH masses consistent with scenario pGas ($M_\bullet \sim 10^5
M_\odot$).

Note that, in figure~\ref{fig:bhcomp}, the cut-off is caused by the fact that
even the case for $M_\bullet = 0$ (i.e. the unmodified NFW profile) is ruled
out at $\langle\sigma v\rangle=3\times10^{-26} \text{cm}^3 \text{s}^{-1}$ for
$m_\chi\lesssim40\;\text{GeV}$ for the $\overline{b}b$ channel, as can be seen
in figure~\ref{fig:msigma}. In contrast, the horizontal asymptotic behavior in
figure~\ref{fig:bhcompalt} is caused by the flattening of $\langle\sigma
v\rangle$--$m_\chi$ at low WIMP mass (see also figure~\ref{fig:msigma}). 

\medskip

\begin{figure}[!tbp]
  \centering
  \includegraphics[angle=270,width=0.7\textwidth]{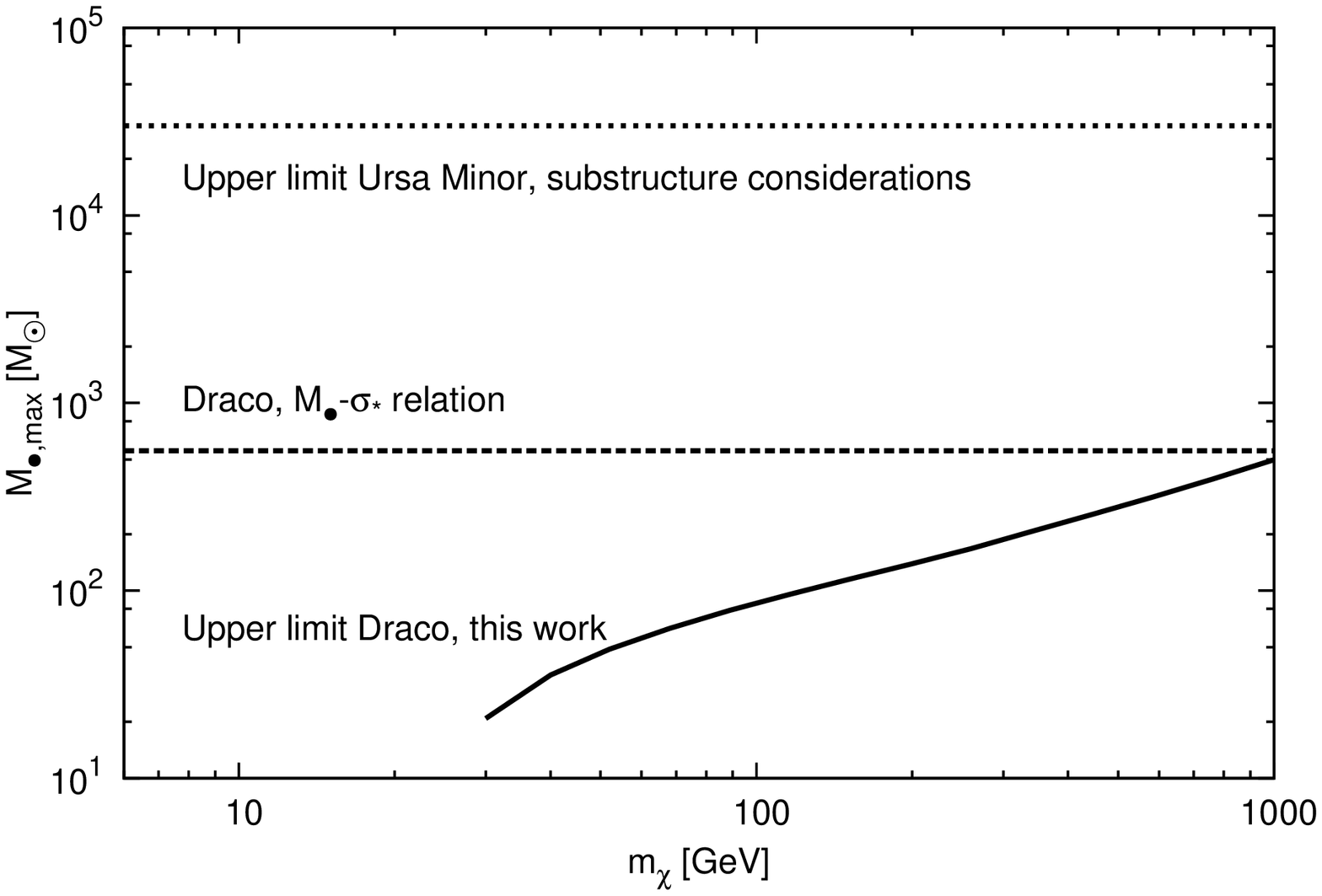}\\
  \caption{Upper limit on the mass of the IMBH for Draco derived in this work
  from annihilation into $\overline{b}b$, compared to the IMBH mass expected from 
  the $M_\bullet$--$\sigma_*$ relation.}
  \label{fig:concl}
\end{figure}

Our upper limits on $M_\bullet$ that we obtained from Fermi LAT observations
and the WIMP hypothesis are stronger than those that we derived from the
methods described above in section~\ref{sec:msigma}, as can be seen in
figure~\ref{fig:concl}. 

Recently, the mini-spike scenario was used along with IMBH 
masses derived from, amongst others, the $M_\bullet$--$\sigma_*$ relation to derive constraints on 
WIMPs~\citep{gonzalez-morales2014}, and these results are consistent with ours. 
From our results it can be seen that the thermal WIMP scenario rules out IMBH masses 
derived from the $M_\bullet$--$\sigma_*$ relation for Draco, for WIMP masses \textless 1 TeV.

We note that in the case of either non-adiabatic
growth or off-center formation (see section~\ref{sec:persist}), our constraints
become much looser; $M_\bullet \lesssim 10^7$--$10^8 M_\odot$.

\subsection{Other dwarf spheroidal galaxies}

\begin{figure}[!tbp]
    \caption{Upper limits on gamma-ray flux (\ref{fig:limitsursas}) and black
    hole mass $M_\bullet$ (\ref{fig:bhupursas}) for Ursa Minor and Ursa Major
    II as a function of WIMP mass $m_\chi$, in comparison to Draco, for the
    $\overline{b}b$ channel and a thermal cross section.}
    \label{fig:ursas}
    \centering
    \begin{subfigure}[!htbp]{0.45\textwidth}
    \includegraphics[angle=270, width=\textwidth]{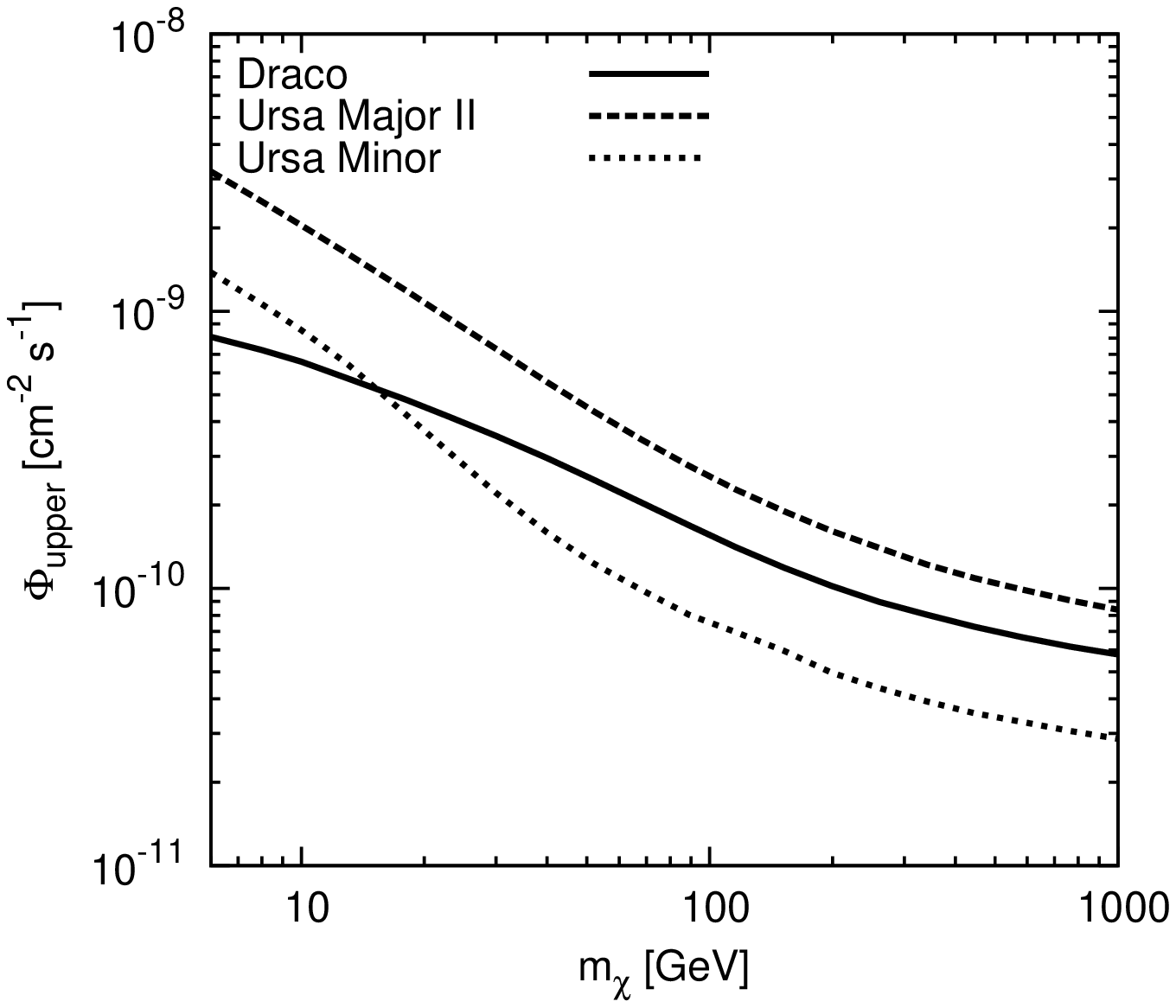}
      \caption{} \label{fig:limitsursas}
    \end{subfigure}\;
     \begin{subfigure}[!htbp]{0.45\textwidth}
    \includegraphics[angle=270, width=\textwidth]{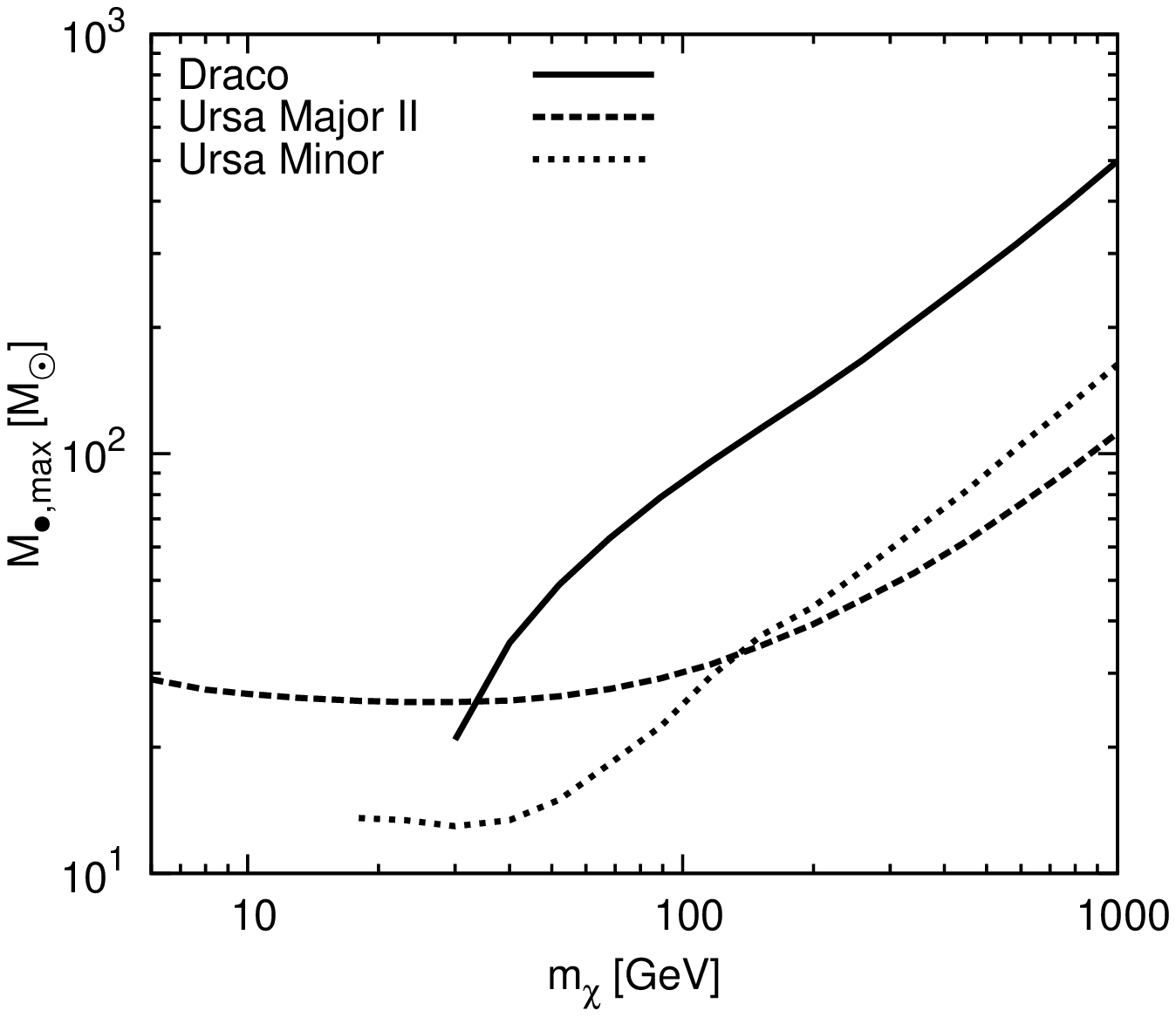}
       \caption{}\label{fig:bhupursas}
    \end{subfigure}
\end{figure}

Considering heliocentric distance and expected J-factor, the two next best
candidates after Draco for detection of gamma-rays from DM annihilation appear
to be the Ursa Minor and Ursa Major II dSphs (see table~\ref{tab:dsph}). To
check for consistency, we repeat our analysis for these objects.
%
%
Our results for $\Phi_{\text{upper}}$ and $M_{\bullet, \text{max}}$ in the
thermal relic and $t_\bullet = 10^{10} \;\text{yr}$ case for Ursa Minor and
Ursa Major II are shown in figure~\ref{fig:ursas}, and are comparable to the
results previously obtained for Draco. 

\medskip

We can use known upper limits on the gamma-ray flux from nearly \emph{all}
Galactic dSphs (specifically, 18 dSphs), as derived in ref.~\cite{fermi2013}, to infer
general limits on IMBHs in scenario pGas. As mentioned in
section~\ref{sec:fluxlims}, the presence of a mini-spike originating from the
formation of a scenario pGas IMBH would cause the expected gamma-ray flux from
Draco to overshoot the upper limits by 2 to 3 orders
of magnitude.  Given that from both, our comparison of Ursa Minor and Ursa
Major II to Draco as well as already known results~\citep{fermi2013}, the flux
upper limits for the various Galactic dSph are at the same order of magnitude,
and given that the expected flux from a mini-spike does not depend strongly on
the parameters of the original NFW profile of the dSph (see
eq.~\eqref{eq:phiscale}), we conclude that the existence of scenario pGas IMBHs
without previous BH-BH merger is ruled out for all Galactic dSphs mentioned in
ref.~\citep{fermi2013}.

We can extend this argument to dSphs in M31, though there are currently no
upper limits on the gamma-ray flux from these objects. In
section~\ref{sec:MWM31}, we selected a population of 12 M31 dSphs with an
average heliocentric distance of 711 kpc. Since the measured flux scales with
the inverse of distance squared, we can rescale the upper limits derived for
e.g.~Draco to the case of an M31 dSph: $76^2/711^2 \simeq 0.01$, meaning that
the expected boost in flux of 2--3 orders of magnitude would still be visible
in M31 dSphs.  It is hence plausible that also M31 dSphs do not harbor any
unmerged IMBH from scenario pGas.

Finally, we remind that in our above Monte Carlo simulations of the BH merger
history we inferred that in the case of scenario pGas the chances of none of the
MW and M31 dSphs harboring an unmerged IMBH is 16\%.  We hence find that--if
DM is made out of WIMPs with typical masses and cross sections--scenario pGas
is in slight tension with the existing gamma-ray data at about 84\% CL.

\section{Conclusions}
\label{sec:conclusion}

We have derived constraints on the existence of IMBHs in dwarf spheroidal galaxies, 
focusing in particular on the case where DM is made of WIMPs with typical masses and self-annihilation
cross sections.  We focused in particular on a formation scenario, pGas, where IMBH
originate from massive objects formed directly during the collapse of
primordial gas in early-forming halos (with $M_\bullet \sim 10^5 M_\odot$).
IMBH generated by the collapse of Pop III stars are unlikely to feature
surviving mini-spikes. 

Using almost 6 years of \Fermi\ LAT data we determined an upper limit on the DM
annihilation flux (figure~\ref{fig:lims}) and cross section
(figure~\ref{fig:msigma}) as a function of WIMP mass for five different
annihilation channels from dSphs Draco, Ursa Minor and Ursa Major II, which we
modeled as point sources with a DM distribution following a spiked NFW profile.
With these flux upper limits we were then able to derive upper limits on the
mass of a central IMBH in the mini-spike scenario for different annihilation
cross sections (figure~\ref{fig:bhcomps}).  

Our conclusions are as follows: 

\begin{itemize}
    \item Under the assumption that the IMBH grew adiabatically at the center
        of the dSph, we effectively rule out the existence of thermal WIMPs mini-spikes 
        for IMBH masses $M_\bullet \gtrsim
        10^3 M_\odot$ for the Draco,
        Ursa Minor and Ursa Major II dSphs with high significance
        (figures~\ref{fig:bhcomp},~\ref{fig:bhupursas}).
    \item  Our constraints on the IMBH mass $M_\bullet$ are stronger than
        previous constraints derived with other methods, e.g.~the $M_\bullet -
        \sigma_*$ relation or substructure considerations
        (figure~\ref{fig:concl})
    \item Our Monte Carlo simulations of merger histories of BH in dSphs in the pGas IMBH formation scenario
        indicate that, while the probability of an individual dSph hosting an
        IMBH is quite low (about 10\%), it is somewhat
        unlikely (probability 16\%) that \emph{none} of the dSph satellites of the MW and M31 host
        such an IMBH with a surviving mini-spike.
    \item Given the fact that under the thermal WIMPs hypothesis IMBHs without previous
        merger (and hence a surviving mini-spike) and with a mass of
        $\sim10^5\, M_\odot$ can be plausibly excluded for \emph{all} of the MW
        and M31 dSphs, we find that scenario pGas is disfavoured by
        existing gamma-ray data at about $84\%$ CL.
\end{itemize}
\paragraph{Acknowledgement} We thank S. Profumo and A. X. Gonzalez-Morales for providing additional information on their recent work \cite{gonzalez-morales2014}, which allowed a detailed comparison of our results. GB acknowledges support from the European Research Council through the ERC Starting Grant \emph{WIMPs Kairos}
\bibliography{report}

\providecommand{\href}[2]{#2}\begingroup\raggedright\begin{thebibliography}{10}

\bibitem{peebles1972}
P.~J.~E. {Peebles}, {\it {Star Distribution Near a Collapsed Object}},  {\em
  \apj} {\bf 178} (Dec., 1972) 371--376.

\bibitem{Quinlan:1994ed}
G.~D. Quinlan, L.~Hernquist, and S.~Sigurdsson, {\it {Models of Galaxies with
  Central Black Holes: Adiabatic Growth in Spherical Galaxies}},  {\em
  Astrophys.J.} {\bf 440} (1995) 554--564,
  [\href{http://arxiv.org/abs/astro-ph/9407005}{{\tt astro-ph/9407005}}].

\bibitem{BertoneBook}
G.~{Bertone}, ed., {\em {Particle Dark Matter: Observations, Models and
  Searches}}.
\newblock Cambridge University Press, 2010.

\bibitem{gondolosilk1999}
P.~{Gondolo} and J.~{Silk}, {\it {Dark Matter Annihilation at the Galactic
  Center}},  {\em Physical Review Letters} {\bf 83} (Aug., 1999) 1719--1722,
  [\href{http://arxiv.org/abs/astro-ph/9906391}{{\tt astro-ph/9906391}}].

\bibitem{Jungman:1995df}
G.~Jungman, M.~Kamionkowski, and K.~Griest, {\it {Supersymmetric dark matter}},
   {\em Phys.Rept.} {\bf 267} (1996) 195--373,
  [\href{http://arxiv.org/abs/hep-ph/9506380}{{\tt hep-ph/9506380}}].

\bibitem{Bergstrom:2000pn}
L.~Bergstrom, {\it {Nonbaryonic dark matter: Observational evidence and
  detection methods}},  {\em Rept.Prog.Phys.} {\bf 63} (2000) 793,
  [\href{http://arxiv.org/abs/hep-ph/0002126}{{\tt hep-ph/0002126}}].

\bibitem{bhs2005}
G.~{Bertone}, D.~{Hooper}, and J.~{Silk}, {\it {Particle dark matter: evidence,
  candidates and constraints}},  {\em \physrep} {\bf 405} (Jan., 2005)
  279--390, [\href{http://arxiv.org/abs/hep-ph/0404175}{{\tt hep-ph/0404175}}].

\bibitem{Gondolo:2000pn}
P.~Gondolo, {\it {Either neutralino dark matter or cuspy dark halos}},  {\em
  Phys.Lett.} {\bf B494} (2000) 181--186,
  [\href{http://arxiv.org/abs/hep-ph/0002226}{{\tt hep-ph/0002226}}].

\bibitem{Bertone:2001jv}
G.~Bertone, G.~Sigl, and J.~Silk, {\it {Astrophysical limits on massive dark
  matter}},  {\em Mon.Not.Roy.Astron.Soc.} {\bf 326} (2001) 799--804,
  [\href{http://arxiv.org/abs/astro-ph/0101134}{{\tt astro-ph/0101134}}].

\bibitem{Bertone:2002je}
G.~Bertone, G.~Sigl, and J.~Silk, {\it {Annihilation radiation from a dark
  matter spike at the galactic center}},  {\em Mon.Not.Roy.Astron.Soc.} {\bf
  337} (2002) 98, [\href{http://arxiv.org/abs/astro-ph/0203488}{{\tt
  astro-ph/0203488}}].

\bibitem{Ahn:2007ty}
E.-J. Ahn, G.~Bertone, and D.~Merritt, {\it {Impact of astrophysical processes
  on the gamma-ray background from dark matter annihilations}},  {\em
  Phys.Rev.} {\bf D76} (2007) 023517,
  [\href{http://arxiv.org/abs/astro-ph/0703236}{{\tt astro-ph/0703236}}].

\bibitem{Belikov:2013nca}
A.~Belikov and J.~Silk, {\it {Diffuse Gamma Ray Background from Annihilating
  Dark Matter in Density Spikes around Supermassive Black Holes}},  {\em
  Phys.Rev.} {\bf D89} (2014) 043520,
  [\href{http://arxiv.org/abs/1312.0007}{{\tt arXiv:1312.0007}}].

\bibitem{uzk2001}
P.~{Ullio}, H.~{Zhao}, and M.~{Kamionkowski}, {\it {Dark-matter spike at the
  galactic center?}},  {\em \prd} {\bf 64} (Aug., 2001) 043504,
  [\href{http://arxiv.org/abs/astro-ph/0101481}{{\tt astro-ph/0101481}}].

\bibitem{merritt2002}
D.~{Merritt}, M.~{Milosavljevi{\'c}}, L.~{Verde}, and R.~{Jimenez}, {\it {Dark
  Matter Spikes and Annihilation Radiation from the Galactic Center}},  {\em
  Physical Review Letters} {\bf 88} (May, 2002) 191301,
  [\href{http://arxiv.org/abs/astro-ph/0201376}{{\tt astro-ph/0201376}}].

\bibitem{bertonemerrit2005}
G.~{Bertone} and D.~{Merritt}, {\it {Dark Matter Dynamics and Indirect
  Detection}},  {\em Modern Physics Letters A} {\bf 20} (2005) 1021--1036,
  [\href{http://arxiv.org/abs/astro-ph/0504422}{{\tt astro-ph/0504422}}].

\bibitem{2014Natur.513...74P}
D.~R. {Pasham}, T.~E. {Strohmayer}, and R.~F. {Mushotzky}, {\it {A
  400-solar-mass black hole in the galaxy M82}},  {\em \nat} {\bf 513} (Sept.,
  2014) 74--76.

\bibitem{zhao2005}
H.~{Zhao} and J.~{Silk}, {\it {Dark Minihalos with Intermediate Mass Black
  Holes}},  {\em Physical Review Letters} {\bf 95} (June, 2005) 011301,
  [\href{http://arxiv.org/abs/astro-ph/0501625}{{\tt astro-ph/0501625}}].

\bibitem{Bertone:2006nq}
G.~Bertone, {\it {Prospects for detecting dark matter with neutrino telescopes
  in intermediate mass black holes scenarios}},  {\em Phys.Rev.} {\bf D73}
  (2006) 103519, [\href{http://arxiv.org/abs/astro-ph/0603148}{{\tt
  astro-ph/0603148}}].

\bibitem{Fornasa:2007nr}
M.~Fornasa and G.~Bertone, {\it {Black Holes as Dark Matter Annihilation
  Boosters}},  {\em Int.J.Mod.Phys.} {\bf D17} (2008) 1125--1157,
  [\href{http://arxiv.org/abs/0711.3148}{{\tt arXiv:0711.3148}}].

\bibitem{Aharonian:2008wt}
{\bf HESS Collaboration} Collaboration, F.~Aharonian et~al., {\it {Search for
  Gamma-rays from Dark Matter annihilations around Intermediate Mass Black
  Holes with the H.E.S.S. experiment}},  {\em Phys.Rev.} {\bf D78} (2008)
  072008, [\href{http://arxiv.org/abs/0806.2981}{{\tt arXiv:0806.2981}}].

\bibitem{Taoso:2008qz}
M.~Taoso, S.~Ando, G.~Bertone, and S.~Profumo, {\it {Angular correlations in
  the cosmic gamma-ray background from dark matter annihilation around
  intermediate-mass black holes}},  {\em Phys.Rev.} {\bf D79} (2009) 043521,
  [\href{http://arxiv.org/abs/0811.4493}{{\tt arXiv:0811.4493}}].

\bibitem{Bringmann:2009ip}
T.~Bringmann, J.~Lavalle, and P.~Salati, {\it {Intermediate Mass Black Holes
  and Nearby Dark Matter Point Sources: A Myth-Buster}},  {\em Phys.Rev.Lett.}
  {\bf 103} (2009) 161301, [\href{http://arxiv.org/abs/0902.3665}{{\tt
  arXiv:0902.3665}}].

\bibitem{Bertone:2009kj}
G.~Bertone, M.~Fornasa, M.~Taoso, and A.~R. Zentner, {\it {Dark Matter
  Annihilation around Intermediate Mass Black Holes: an update}},  {\em New
  J.Phys.} {\bf 11} (2009) 105016, [\href{http://arxiv.org/abs/0905.4736}{{\tt
  arXiv:0905.4736}}].

\bibitem{safonova2010}
M.~{Safonova} and P.~{Shastri}, {\it {Extrapolating SMBH correlations down the
  mass scale: the case for IMBHs in globular clusters}},  {\em \apss} {\bf 325}
  (Jan., 2010) 47--58, [\href{http://arxiv.org/abs/0910.2551}{{\tt
  arXiv:0910.2551}}].

\bibitem{collins2014}
M.~L.~M. {Collins}, S.~C. {Chapman}, R.~M. {Rich}, R.~A. {Ibata}, N.~F.
  {Martin}, M.~J. {Irwin}, N.~F. {Bate}, G.~F. {Lewis}, J.~{Pe{\~n}arrubia},
  N.~{Arimoto}, C.~M. {Casey}, A.~M.~N. {Ferguson}, A.~{Koch}, A.~W.
  {McConnachie}, and N.~{Tanvir}, {\it {The Masses of Local Group Dwarf
  Spheroidal Galaxies: The Death of the Universal Mass Profile}},  {\em \apj}
  {\bf 783} (Mar., 2014) 7, [\href{http://arxiv.org/abs/1309.3053}{{\tt
  arXiv:1309.3053}}].

\bibitem{abdo2010}
{\bf Fermi LAT} Collaboration, A.~A. {Abdo}, M.~{Ackermann}, M.~{Ajello}, W.~B.
  {Atwood}, L.~{Baldini}, J.~{Ballet}, G.~{Barbiellini}, D.~{Bastieri},
  K.~{Bechtol}, and et~al., {\it {Observations of Milky Way Dwarf Spheroidal
  Galaxies with the Fermi-Large Area Telescope Detector and Constraints on Dark
  Matter Models}},  {\em \apj} {\bf 712} (Mar., 2010) 147--158,
  [\href{http://arxiv.org/abs/1001.4531}{{\tt arXiv:1001.4531}}].

\bibitem{mateo1998}
M.~L. {Mateo}, {\it {Dwarf Galaxies of the Local Group}},  {\em \araa} {\bf 36}
  (1998) 435--506, [\href{http://arxiv.org/abs/astro-ph/9810070}{{\tt
  astro-ph/9810070}}].

\bibitem{gallagher2003}
J.~S. {Gallagher}, G.~J. {Madsen}, R.~J. {Reynolds}, E.~K. {Grebel}, and T.~A.
  {Smecker-Hane}, {\it {A Search for Ionized Gas in the Draco and Ursa Minor
  Dwarf Spheroidal Galaxies}},  {\em \apj} {\bf 588} (May, 2003) 326--330,
  [\href{http://arxiv.org/abs/astro-ph/0301228}{{\tt astro-ph/0301228}}].

\bibitem{grcevich2009}
J.~{Grcevich} and M.~E. {Putman}, {\it {H I in Local Group Dwarf Galaxies and
  Stripping by the Galactic Halo}},  {\em \apj} {\bf 696} (May, 2009) 385--395,
  [\href{http://arxiv.org/abs/0901.4975}{{\tt arXiv:0901.4975}}].

\bibitem{segall2007}
M.~{S{\'e}gall}, R.~A. {Ibata}, M.~J. {Irwin}, N.~F. {Martin}, and
  S.~{Chapman}, {\it {Draco, a flawless dwarf galaxy}},  {\em \mnras} {\bf 375}
  (Mar., 2007) 831--842, [\href{http://arxiv.org/abs/astro-ph/0612263}{{\tt
  astro-ph/0612263}}].

\bibitem{nfw1997}
J.~F. {Navarro}, C.~S. {Frenk}, and S.~D.~M. {White}, {\it {A Universal Density
  Profile from Hierarchical Clustering}},  {\em \apj} {\bf 490} (Dec., 1997)
  493, [\href{http://arxiv.org/abs/astro-ph/9611107}{{\tt astro-ph/9611107}}].

\bibitem{swartz2004}
D.~A. {Swartz}, K.~K. {Ghosh}, A.~F. {Tennant}, and K.~{Wu}, {\it {The
  Ultraluminous X-Ray Source Population from the Chandra Archive of Galaxies}},
   {\em \apjs} {\bf 154} (Oct., 2004) 519--539,
  [\href{http://arxiv.org/abs/astro-ph/0405498}{{\tt astro-ph/0405498}}].

\bibitem{volonteri2010}
M.~{Volonteri}, {\it {Formation of supermassive black holes}},  {\em \aapr}
  {\bf 18} (July, 2010) 279--315, [\href{http://arxiv.org/abs/1003.4404}{{\tt
  arXiv:1003.4404}}].

\bibitem{bzs2005}
G.~{Bertone}, A.~R. {Zentner}, and J.~{Silk}, {\it {New signature of dark
  matter annihilations: Gamma rays from intermediate-mass black holes}},  {\em
  \prd} {\bf 72} (Nov., 2005) 103517,
  [\href{http://arxiv.org/abs/astro-ph/0509565}{{\tt astro-ph/0509565}}].

\bibitem{madau2001}
P.~{Madau} and M.~J. {Rees}, {\it {Massive Black Holes as Population III
  Remnants}},  {\em \apjl} {\bf 551} (Apr., 2001) L27--L30,
  [\href{http://arxiv.org/abs/astro-ph/0101223}{{\tt astro-ph/0101223}}].

\bibitem{heger2003}
A.~{Heger}, C.~L. {Fryer}, S.~E. {Woosley}, N.~{Langer}, and D.~H. {Hartmann},
  {\it {How Massive Single Stars End Their Life}},  {\em \apj} {\bf 591} (July,
  2003) 288--300, [\href{http://arxiv.org/abs/astro-ph/0212469}{{\tt
  astro-ph/0212469}}].

\bibitem{koushiappas2004}
S.~M. {Koushiappas}, J.~S. {Bullock}, and A.~{Dekel}, {\it {Massive black hole
  seeds from low angular momentum material}},  {\em \mnras} {\bf 354} (Oct.,
  2004) 292--304, [\href{http://arxiv.org/abs/astro-ph/0311487}{{\tt
  astro-ph/0311487}}].

\bibitem{shapiro1983}
S.~L. {Shapiro}, S.~A. {Teukolsky}, and A.~P. {Lightman}, {\it {Black Holes,
  White Dwarfs, and Neutron Stars: The Physics of Compact Objects}},  {\em
  Physics Today} {\bf 36} (1983) 89.

\bibitem{mashchenko2006}
S.~{Mashchenko}, A.~{Sills}, and H.~M. {Couchman}, {\it {Constraining Global
  Properties of the Draco Dwarf Spheroidal Galaxy}},  {\em \apj} {\bf 640}
  (Mar., 2006) 252--269, [\href{http://arxiv.org/abs/astro-ph/0511567}{{\tt
  astro-ph/0511567}}].

\bibitem{merritt2004}
D.~{Merritt}, {\it {Single and Binary Black Holes and their Influence on
  Nuclear Structure}},  {\em Coevolution of Black Holes and Galaxies} (2004)
  263, [\href{http://arxiv.org/abs/astro-ph/0301257}{{\tt astro-ph/0301257}}].

\bibitem{castellani1975}
V.~{Castellani}, {\it {On the age of the Draco dwarf galaxy}},  {\em \mnras}
  {\bf 172} (Sept., 1975) 59P--64P.

\bibitem{2007PhRvD..76j3532V}
E.~{Vasiliev}, {\it {Dark matter annihilation near a black hole: Plateau versus
  weak cusp}},  {\em \prd} {\bf 76} (Nov., 2007) 103532,
  [\href{http://arxiv.org/abs/0707.3334}{{\tt arXiv:0707.3334}}].

\bibitem{Fornasa:2007ap}
M.~Fornasa, M.~Taoso, and G.~Bertone, {\it {Gamma-Rays from Dark Matter
  Mini-Spikes in M31}},  {\em Phys.Rev.} {\bf D76} (2007) 043517,
  [\href{http://arxiv.org/abs/astro-ph/0703757}{{\tt astro-ph/0703757}}].

\bibitem{bertone2009}
G.~{Bertone}, M.~{Fornasa}, M.~{Taoso}, and A.~R. {Zentner}, {\it {Dark matter
  annihilation around intermediate mass black holes: an update}},  {\em New
  Journal of Physics} {\bf 11} (Oct., 2009) 105016,
  [\href{http://arxiv.org/abs/0905.4736}{{\tt arXiv:0905.4736}}].

\bibitem{2014arXiv1406.4856F}
B.~D. {Fields}, S.~L. {Shapiro}, and J.~{Shelton}, {\it {Galactic Center
  Gamma-Ray Excess from Dark Matter Annihilation: Is There A Black Hole
  Spike?}},  {\em ArXiv e-prints} (June, 2014)
  [\href{http://arxiv.org/abs/1406.4856}{{\tt arXiv:1406.4856}}].

\bibitem{2010MNRAS.408.1139V}
S.~{van Wassenhove}, M.~{Volonteri}, M.~G. {Walker}, and J.~R. {Gair}, {\it
  {Massive black holes lurking in Milky Way satellites}},  {\em \mnras} {\bf
  408} (Oct., 2010) 1139--1146, [\href{http://arxiv.org/abs/1001.5451}{{\tt
  arXiv:1001.5451}}].

\bibitem{VHM}
M.~{Volonteri}, F.~{Haardt}, and P.~{Madau}, {\it {The Assembly and Merging
  History of Supermassive Black Holes in Hierarchical Models of Galaxy
  Formation}},  {\em \apj} {\bf 582} (Jan., 2003) 559--573.

\bibitem{Yoshida2006}
N.~{Yoshida}, K.~{Omukai}, L.~{Hernquist}, and T.~{Abel}, {\it {Formation of
  Primordial Stars in a {$\Lambda$}CDM Universe}},  {\em \apj} {\bf 652} (Nov.,
  2006) 6--25, [\href{http://arxiv.org/abs/astro-ph/}{{\tt astro-ph/}}].

\bibitem{Scannapieco2003}
E.~{Scannapieco}, R.~{Schneider}, and A.~{Ferrara}, {\it {The Detectability of
  the First Stars and Their Cluster Enrichment Signatures}},  {\em ApJ} {\bf
  589} (May, 2003) 35--52, [\href{http://arxiv.org/abs/astro-ph/0301628}{{\tt
  astro-ph/0301628}}].

\bibitem{LN2006}
G.~{Lodato} and P.~{Natarajan}, {\it {Supermassive black hole formation during
  the assembly of pre-galactic discs}},  {\em \mnras} {\bf 371} (Oct., 2006)
  1813--1823, [\href{http://arxiv.org/abs/astro-ph/}{{\tt astro-ph/}}].

\bibitem{VLN2008}
M.~{Volonteri}, G.~{Lodato}, and P.~{Natarajan}, {\it {The evolution of massive
  black hole seeds}},  {\em MNRAS} {\bf 383} (Jan., 2008) 1079--1088,
  [\href{http://arxiv.org/abs/0709.0529}{{\tt 0709.0529}}].

\bibitem{MoMaoWhite1998}
H.~J. {Mo}, S.~{Mao}, and S.~D.~M. {White}, {\it {The formation of galactic
  discs}},  {\em {MNRAS}} {\bf 295} (Apr., 1998) 319--336.

\bibitem{Binney1987}
J.~{Binney} and S.~{Tremaine}, {\em {Galactic Dynamics: Second Edition}}.
\newblock Galactic Dynamics: Second Edition, by James Binney and Scott
  Tremaine.~ISBN 978-0-691-13026-2 (HB).~Published by Princeton University
  Press, Princeton, NJ USA, 2008., 2008.

\bibitem{Taylor2001}
J.~E. {Taylor} and A.~{Babul}, {\it {The Dynamics of Sinking Satellites around
  Disk Galaxies: A Poor Man's Alternative to High-Resolution Numerical
  Simulations}},  {\em ApJ} {\bf 559} (Oct., 2001) 716--735,
  [\href{http://arxiv.org/abs/astro-ph/}{{\tt astro-ph/}}].

\bibitem{fermi2013}
{\bf Fermi LAT} Collaboration, M.~{Ackermann}, A.~{Albert}, B.~{Anderson},
  L.~{Baldini}, and et~al., {\it {Dark Matter Constraints from Observations of
  25 Milky Way Satellite Galaxies with the Fermi Large Area Telescope}},  {\em
  ArXiv e-prints} (Oct., 2013) [\href{http://arxiv.org/abs/1310.0828}{{\tt
  arXiv:1310.0828}}].

\bibitem{splash2012}
E.~J. {Tollerud}, R.~L. {Beaton}, M.~C. {Geha}, and et~al., {\it {The SPLASH
  Survey: Spectroscopy of 15 M31 Dwarf Spheroidal Satellite Galaxies}},  {\em
  \apj} {\bf 752} (June, 2012) 45, [\href{http://arxiv.org/abs/1112.1067}{{\tt
  arXiv:1112.1067}}].

\bibitem{wolf2010}
J.~{Wolf}, G.~D. {Martinez}, J.~S. {Bullock}, M.~{Kaplinghat}, M.~{Geha}, R.~R.
  {Mu{\~n}oz}, J.~D. {Simon}, and F.~F. {Avedo}, {\it {Accurate masses for
  dispersion-supported galaxies}},  {\em \mnras} {\bf 406} (Aug., 2010)
  1220--1237, [\href{http://arxiv.org/abs/0908.2995}{{\tt arXiv:0908.2995}}].

\bibitem{tremaine2002}
S.~{Tremaine}, K.~{Gebhardt}, R.~{Bender}, G.~{Bower}, A.~{Dressler}, S.~M.
  {Faber}, A.~V. {Filippenko}, R.~{Green}, C.~{Grillmair}, L.~C. {Ho},
  J.~{Kormendy}, T.~R. {Lauer}, J.~{Magorrian}, J.~{Pinkney}, and
  D.~{Richstone}, {\it {The Slope of the Black Hole Mass versus Velocity
  Dispersion Correlation}},  {\em \apj} {\bf 574} (Aug., 2002) 740--753,
  [\href{http://arxiv.org/abs/astro-ph/0203468}{{\tt astro-ph/0203468}}].

\bibitem{lora2009}
V.~{Lora}, F.~J. {S{\'a}nchez-Salcedo}, A.~C. {Raga}, and A.~{Esquivel}, {\it
  {An Upper Limit on the Mass of the Black Hole in Ursa Minor Dwarf Galaxy}},
  {\em \apjl} {\bf 699} (July, 2009) L113--L117,
  [\href{http://arxiv.org/abs/0906.0951}{{\tt arXiv:0906.0951}}].

\bibitem{Ferrarese2005}
L.~{Ferrarese} and H.~{Ford}, {\it {Supermassive Black Holes in Galactic
  Nuclei: Past, Present and Future Research}},  {\em Space Science Reviews}
  {\bf 116} (Feb., 2005) 523--624,
  [\href{http://arxiv.org/abs/astro-ph/0411247}{{\tt astro-ph/0411247}}].

\bibitem{2013ARA&A..51..511K}
J.~{Kormendy} and L.~C. {Ho}, {\it {Coevolution (Or Not) of Supermassive Black
  Holes and Host Galaxies}},  {\em \araa} {\bf 51} (Aug., 2013) 511--653,
  [\href{http://arxiv.org/abs/1304.7762}{{\tt arXiv:1304.7762}}].

\bibitem{2012ApJ...746...89J}
J.~R. {Jardel} and K.~{Gebhardt}, {\it {The Dark Matter Density Profile of the
  Fornax Dwarf}},  {\em \apj} {\bf 746} (Feb., 2012) 89,
  [\href{http://arxiv.org/abs/1112.0319}{{\tt arXiv:1112.0319}}].

\bibitem{2007ApJ...667L..53W}
M.~G. {Walker}, M.~{Mateo}, E.~W. {Olszewski}, O.~Y. {Gnedin}, X.~{Wang},
  B.~{Sen}, and M.~{Woodroofe}, {\it {Velocity Dispersion Profiles of Seven
  Dwarf Spheroidal Galaxies}},  {\em \apjl} {\bf 667} (Sept., 2007) L53--L56,
  [\href{http://arxiv.org/abs/0708.0010}{{\tt arXiv:0708.0010}}].

\bibitem{1982MNRAS.200..361B}
J.~{Binney} and G.~A. {Mamon}, {\it {M/L and velocity anisotropy from
  observations of spherical galaxies, or must M87 have a massive black hole}},
  {\em \mnras} {\bf 200} (July, 1982) 361--375.

\bibitem{2003MNRAS.343..401L}
E.~L. {{\L}okas} and G.~A. {Mamon}, {\it {Dark matter distribution in the Coma
  cluster from galaxy kinematics: breaking the mass-anisotropy degeneracy}},
  {\em \mnras} {\bf 343} (Aug., 2003) 401--412,
  [\href{http://arxiv.org/abs/astro-ph/0302461}{{\tt astro-ph/0302461}}].

\bibitem{fermi2010}
{\bf Fermi LAT} Collaboration, A.~A. Abdo, M.~Ackermann, M.~Ajello, and et~al.,
  {\it Spectrum of the isotropic diffuse gamma-ray emission derived from
  first-year fermi large area telescope data},  {\em Phys. Rev. Lett.} {\bf
  104} (Mar, 2010) 101101.

\bibitem{pppc}
M.~{Cirelli}, G.~{Corcella}, A.~{Hektor}, G.~{H{\"u}tsi}, M.~{Kadastik},
  P.~{Panci}, M.~{Raidal}, F.~{Sala}, and A.~{Strumia}, {\it {PPPC 4 DM ID: a
  poor particle physicist cookbook for dark matter indirect detection}},  {\em
  \jcap} {\bf 3} (Mar., 2011) 51, [\href{http://arxiv.org/abs/1012.4515}{{\tt
  arXiv:1012.4515}}].

\bibitem{gonzalez-morales2014}
A.~X. {Gonzalez-Morales}, S.~{Profumo}, and F.~S. {Queiroz}, {\it {The Effect
  of Black Holes in Local Dwarf Spheroidal Galaxies on Gamma-Ray Constraints on
  Dark Matter Annihilation}},  {\em ArXiv e-prints} (June, 2014)
  [\href{http://arxiv.org/abs/1406.2424}{{\tt arXiv:1406.2424}}].

\end{thebibliography}\endgroup

\end{document}